# Realspace Sovereigns in Cyberspace: The Case of Domain Names

Catherine T. Struve & R. Polk Wagner
*cstruve@law.upenn.edu*  *polk@law.upenn.edu*

*Version 2.1: October 25, 2001*

**DRAFT**
PLEASE CHECK FOR NEW
VERSION PRIOR TO CITATION

University of Pennsylvania Law School
3400 Chestnut Street
Philadelphia, Pennsylvania 19104

# Realspace Sovereigns in Cyberspace: The Case of Domain Names
*R. Polk Wagner and Catherine T. Struve*

## Abstract

In the era of the global Internet, realspace sovereigns face new problems relating to the scope and enforceability of their laws, many of which are intended to protect local individuals and commercial entities. How these traditional sovereigns respond to these challenges will have far-reaching implications for the ordering of social and commercial behaviour online. In this piece, we take up the case of the domain name system as an example of challenges and solutions yet to come.

We begin by critically analyzing the United States' response to international disputes over domain names, the Anticybersquatting Consumer Protection Act (ACPA), which offers a number of potential remedies to U.S. trademark owners whose marks are registered as domain names by alleged cybersquatters. Most important for our purposes, 15 U.S.C. § 1125(d)(2) permits a trademark owner to seek cancellation or transfer of the domain name by proceeding *in rem* against the domain name itself, thereby purporting to expand the scope of the ACPA to encompass disputes with little direct connection with the United States. Congress appears to have developed § 1125(d)(2) based on a misunderstanding of the constitutional requirements for adjudicative jurisdiction in the U.S. courts; and early court decisions interpreting the provision have perpetuated the misunderstanding. In reality, there exist no cases of foreign cybersquatting (aside from certain cases involving anonymous registrants) as to which the *in rem* provision will be both applicable and constitutional. Thus, we conclude that the provision is notable mainly for its aggressive approach to jurisdiction. This assessment leads us to a broader consideration of the role of realspace sovereigns (especially the United States) in the international regulation of the domain names system.

Turning to this issue, we find that realspace sovereigns have a critical (and yet overlooked) role in the continued viability of a global unsegmented domain names system. By mapping the logical control over the domain names system – the distributed hierarchy that is the basis of the system's design – onto realspace territory, the potential for sovereign regulation of the system becomes apparent. Such regulation can either be supported by recognized principles of prescriptive jurisdiction in international law, or be the de facto result of the geographic facts of the domain names system. We argue that the regulatory significance of geography, and the essentially arbitrary nature of the present territorial locations of the key components of the domain name system, lead to incentives for realspace sovereigns to alter the geographic facts – primarily by establishing alternative root server systems. This in turn implies the future segmentation of the domain name system, and the resulting dramatic decrease in its value. Accordingly, we argue that realspace sovereigns (and especially the United States) have strong interests in avoiding segmentation, and thus must seek to coordinate the regulation of the system.

**Realspace Sovereigns in Cyberspace: The Case of Domain Names**
*Catherine T. Struve and R. Polk Wagner*

# I

In the era of the global Internet, realspace sovereigns are presented with new problems relating to the scope and enforceability of their laws, many of which are intended to protect local individuals and commercial entities. How these traditional sovereigns respond to these challenges will have far-reaching implications for the ordering of social and economic behavior online.[1] In this piece, we take up the case of the domain name system as an example of challenges and solutions yet to come. We begin by critically analyzing the United States' response to international disputes over domain names, especially the Anticybersquatting Consumer Protection Act (ACPA), which offers a number of potential remedies to U.S. trademark owners whose marks are registered as domain names by alleged cybersquatters. The ACPA is notable for its assertions of both adjudicative and prescriptive jurisdiction over foreign registrants; and we take up each of those assertions in turn.

Looking first at the ACPA's provisions with respect to adjudicative jurisdiction, we focus on 15 U.S.C. § 1125(d)(2), which purports to authorize a trademark owner to seek cancellation or transfer of a domain name by proceeding "*in rem*" against the domain name itself in cases where the U.S. courts could not assert personal jurisdiction over the alleged cybersquatter. The text and history of the provision indicate that Congress intended to authorize "*in rem*" proceedings in cases where a foreign registrant's lack of contacts with the United States would render a U.S. court's assertion of personal jurisdiction over the registrant unconstitutional under the Due Process Clause. As we discuss, however, in such cases the registrant's lack of U.S. contacts will render "*in rem*" jurisdiction unconstitutional as well. The ACPA's "*in rem*" provision, therefore, fails effectively to reach the cases Congress appears to have been targeting. Aside from its constitutional deficiencies, however, the "*in rem*" provision is conceptually intriguing because it turns on the assertion that the *res* in question (the domain name) is located within the forum (the United States). The ACPA thus attempts to base "*in rem*" jurisdiction on the premise that a domain name is located in the United States whenever either the dealer or the administrator involved in registering or assigning the domain name is U.S.-based.[2] Because the "*in rem*" provision will in most instances be unconstitutional, the theoretical and practical problems with this assertion are unlikely to gain prominence; but they lead naturally to our consideration of the implications of Congress's assertion of prescriptive jurisdiction over domain name disputes, and, more broadly, the role of realspace sovereigns in domain name regulation.

---

[1] There have been many contributions to this field. *See, e.g.*, David Post & David Johnson, *Law & Borders: The Rise of Law in Cyberspace*, 48 STAN. L. REV. 1367 (1996); Jack Goldsmith, *Against Cyberanarchy*, 65 U. CHI L. REV. 1199 (1998). Margaret Jane Radin & R. Polk Wagner, *The Myth of Private Ordering: Rediscovering Legal Realism in Cyberspace*, 73 CHI.-KENT L. REV. 1295 (1999); David Johnson & David Post, *And How Shall the Net Be Governed? A Meditation on the Relative Virtues of Decentralized, Emergent Law*, in COORDINATING THE INTERNET (B. Kahin and J. Keller eds., 1997); Trotter Hardy, *The Proper Legal Regime for Cyberspace*, 55 U. Pitt L. Rev 993 (1994); Henry R Perrit, Jr. *The Internet is Changing International Law*, 73 Chi-Kent L. Rev. 997 (1998); A Michael Froomkin, *Of Governments and Governance*, 14 Berkeley Law & Technology Journal 617 (1999); A. Michael Froomkin, *The Internet as a Source of Regulatory Arbitrage* (book chapter) in BORDERS IN CYBERSPACE (Brian Kahin and Charles Nesson, eds.) (MIT Press, 1997); Jonathan Weinberg, *Internet Governance, in Transnational Cyberspace Law* (Makoto Ibusuki ed. 2000); Jonathan Weinberg, *ICANN and the Problem of Legitimacy*, 50 Duke L.J. 187 (2000); Milton Mueller, *ICANN and Internet Governance: Sorting through the debris of self-regulation.* INFO 1, 6, 477-500, December 1999.

We begin this second line of analysis by describing the aspects of logical control over the domain names system, noting that the distributed hierarchy in the system's design results correlates control over certain components, especially the root servers, with control over the entire system. By mapping the logical control structure onto the contours of realspace, the potential regulatory authority of realspace sovereigns becomes apparent. This regulatory authority can be grounded in either generally-accepted principles of prescriptive jurisdiction, or the de facto result of the physical location of elements of the domain name system. In the prescriptive jurisdiction case, we note that the location of certain elements – specifically the root or TLD servers – within a sovereign's territory will in almost all cases provide substantial international legal support for the state's assertion of jurisdiction. And even if the sovereign's claim is not formally recognized, we describe how control can be exerted nonetheless. In either of these two cases, the basic point is the same: geography matters.

Importantly, however, while geography matters for purposes of what might be called "territorial" control over the domain names system, from a logical standpoint it is largely irrelevant. Indeed, we note that the geographic facts of the domain name system are uniquely mutable, providing both the means and the incentive for realspace sovereigns to increase their regulatory authority by altering the geographic facts – in our example, by creating and mandating an alternative root server system. This ability to exert regulatory influence, we argue, should be disquieting to the global Internet community. For a predictable result flowing from the creation of alternative root servers is the segmentation of the domain names system, and its concomitant reduction in value.

This insight – the important role that realspace sovereigns play in the regulation of the domain names system – calls for a reconsideration of the present regulatory approaches. In particular, we note that the United States (which arguably stands to lose the most from the segmentation of the domain name system) would be better served in pursuing the goal of international coordination concerning domain names regulation, rather than the extensive assertion of jurisdiction found in the ACPA. We conclude by offering some suggestions and observations about the steps that the United States in particular, and realspace sovereigns more generally, might take to effectuate a policy that reflects the substantial interest in an unsegmented domain names

---

[2] The terms used by the ACPA are "registrar" and "registry." To help distinguish the functions of these two entities, we use the terms "dealer" and "administrator." *See infra* text accompanying note _.

system.

## II

## Jurisdictional Overreaching: The Anticybersquatting Consumer Protection Act

"Cybersquatting"–the phenomenon targeted by the ACPA–occurs when a person registers as a domain name a word or phrase trademarked by another, and does so either in the hopes of selling the domain name to the trademark holder, or with the goal of earning advertising revenue from the visits of web users who are looking for the trademark holder's web page. In passing the ACPA, Congress addressed this substantive problem by prohibiting the bad-faith registration of a domain name consisting of another's mark. Noting, however, the difficulties of suing foreign or anonymous domain name registrants, Congress also created an unusual procedural device for use in cases where the registrant cannot be located or subjected to the jurisdiction of a United States court. In such cases, the ACPA authorizes the mark holder to bring an *in rem* suit[3] directly against the domain name itself.[4] Although the available legislative history indicates

---

[3] In an *in personam* action, jurisdiction flows from the court's authority over the defendant's "person," and any resulting judgment is potentially enforceable against any assets of the defendant, wherever located. By contrast, in an *in rem* action of the type authorized by the ACPA, jurisdiction is based on the court's authority over the *res*–here, the domain name–rather than on authority over the defendant's person. Thus, any judgment in an *in rem* action is limited to the value of the *res*, and the judgment can be enforced only against the *res* and not against any other interests of the defendant.

[4] *See* 15 U.S.C. § 1125(d)(2)(A). The ACPA purports to authorize "an *in rem* civil action." *Id.* Technically, the suit authorized by the ACPA could more accurately be termed a *quasi in rem* Type 1 action, because it does not determine the plaintiff's rights in the *res* as against all the world, but rather determines the relative rights of the plaintiff and another specified person (here, the registrant) in the *res*. *See* Hanson v. Denckla, 357 U.S. 235, 246 n.12 (1958).

One district court has recently taken a different view, arguing that "ACPA *in rem* actions . . . are of the 'true *in rem*' genre because they involve the rights of a disputed mark for every potential rights holder." Cable News Network L.P., L.L.L.P. v. Cnnews.com, 2001 WL 1111193, at *4 (E.D. Va. Sept. 18, 2001). This assessment seems at odds with the structure of the ACPA's *in rem* provisions. The notice requirements set forth in those provisions focus on the domain name registrant, and no one else: they require that the plaintiff send notice of the suit to the registrant and "publish[] notice of the action as the court may direct"—measures which would satisfy the due process requirements for notice of suit with respect to the registrant, but not necessarily with respect to other entities that might have claims to the domain name. Similarly, ACPA claims turn on the conduct of the registrant, rather than on the relative rights of the plaintiff and any person other than the registrant. Moreover, the ACPA provides that a successful *in rem* plaintiff may obtain forfeiture, cancellation or transfer of the domain name; but the statute does not suggest that a successful ACPA plaintiff is thereby immunized from claims by any other person asserting a superior right to the domain name.

In any event, the distinction between *in rem* actions and *quasi in rem* Type 1 actions does not affect our analysis of the ACPA's provisions. *Cf.* Restatement Second of Judgments, § 6 cmt. b (questioning "whether the traditional distinction is useful for that Congress believed the *in rem* provision closed a gap in the enforcement tools available to mark holders, the reality is that this provision adds little to the preexisting jurisdictional bases for ACPA suits. The analysis that follows examines the jurisdictional significance of the *in rem* provision, and concludes that its greatest distinction lies not in its utility (which is minimal, due to constitutional problems) but rather in its approach to the location of domain names.

*A*

*The mechanics of the ACPA*

To prevail on a claim under the ACPA, a plaintiff must show that it is the owner of a protected mark, and that the defendant registered, trafficked in or used a domain name that is identical or (in some cases) confusingly similar to the plaintiff's mark.[5] The ACPA also requires the plaintiff to establish that the defendant acted with a "bad faith intent to profit from th[e] mark," and the Act includes a non-exhaustive list of nine factors to be considered by the court when assessing the element of bad faith.[6] Finally, the Act provides a "safe harbor" for registrants who "believed and had reasonable grounds to believe that the use of the domain name was a fair use or otherwise lawful."[7] Where the prohibited acts occurred prior to the Act's passage, the only remedies available are forfeiture or cancellation of the domain name, or transfer of the domain name to the mark owner.[8] For subsequent violations, the Act authorizes the award of damages and costs,[9] and it permits the plaintiff to elect statutory damages of $1,000 to $100,000, as determined by the court.[10] If appropriate, the court may award treble damages, and in exceptional cases the court may also award a reasonable attorney's fee.[11]

---

any purpose"). Accordingly, for simplicity we will use the term "*in rem*" to describe the ACPA's provisions. *Cf.* Shaffer v. Heitner, 433 U.S. 186, 199 n.17 (1977) (for convenience, using "*in rem*" to denote both *in rem* and *quasi in rem*).

[5] *See* 15 U.S.C. § 1125(d)(1)(A).

[6] *Id.* §§ 1125(d)(1)(A)(i), 1125(d)(1)(B).

[7] *Id.* § 1125(d)(1)(B)(ii).

[8] *See* 15 U.S.C. § 1125(d)(1)(C); Anticybersquatting Consumer Protection Act, Pub. L. No. 106-113, § 3010, 113 Stat. 1536 (providing that damages remedy "shall not be available with respect to the registration, trafficking, or use of a domain name that occurs before the date of the enactment of this Act"); Virtual Works, Inc. v. Volkswagen of America, Inc., 239 F.3d 264, 268 (4th Cir. 2001).

[9] *See* 15 U.S.C. § 1117(a) (providing that an ACPA plaintiff may recover "(1) defendant's profits, (2) any damages sustained by the plaintiff, and (3) the costs of the action").

[10] 15 U.S.C. § 1117(d).

[11] *See* 15 U.S.C. § 1117(a); Shields v. Zuccarini, 254 F.3d 476, 487 (3d Cir. 2001) (affirming award of attorney's fees under ACPA).

B

*Anonymous registrants*

The ACPA's drafters believed that the remedies described above would do little good if the plaintiff was unable to discover the registrant's identity. The House Committee report noted that "a significant problem faced by trademark owners in the fight against cybersquatting is the fact that many cybersquatters register domain names under aliases or otherwise provide false information in their registration applications in order to avoid identification and service of process by the mark owner."[12] The federal courts traditionally have disfavored suits against anonymous defendants, and a plaintiff usually must identify and locate the defendant in order to effect service of process.[13] *Columbia Insurance Co. v. Seescandy.com*–a suit initiated prior to the passage of the ACPA–illustrates the problem. The assignee of various trademarks associated with See's Candy Shops, Inc. sued in federal court, asserting federal and state law claims arising from the registration of the domain names "seescandy.com" and "seecandys.com" by "someone other than the plaintiff."[14] Because the registrant had provided incomplete or false information when registering the domain names, the plaintiff was unable "to collect the information necessary to serve the complaint" on the registrant.[15] The district court recognized the plaintiff's need to ascertain the registrant's identity, but–balancing the plaintiff's interests against "the legitimate and valuable right to participate in online forums anonymously or pseudonymously"[16]–it held that the plaintiff must satisfy a four-part test in order to get discovery on the issue.[17]

The ACPA's *in rem* provision addresses this problem by removing the need to identify an evasive registrant. Under § 1125(d)(2)(A)(ii)(II), a mark owner whose rights are violated by a domain name–and who would have had an ACPA claim against the domain name's registrant–may sue the domain name instead of the registrant, if the owner is unable to find the registrant by sending a notice to the postal and email addresses provided by the registrant to the dealer.[18] The Act's requirement that the plaintiff send the notice to the addresses provided by the registrant–coupled with its requirement that the plaintiff publish notice of the action, as directed by the court, promptly after filing suit–satisfy the due process requirements for notice of suit.[19] Thus, in situations where the registrant cannot be identified, the *in rem* provision holds the promise of "provid[ing] meaningful protection to trademark owners while balancing the interests of privacy and anonymity on the Internet."[20]

---

[12] H.R. REP. NO. 106-412 (1999).

[13] *See, e.g.*, Columbia Ins. Co. v. Seescandy.com, 185 F.R.D. 573, 577-78 (N.D. Cal. 1999) (noting "traditional reluctance for permitting filings against John Doe defendants or fictitious names" and stating that "the default requirement in federal court is that the plaintiff must be able to identify the defendant sufficiently that a summons can be served on the defendant").

[14] 185 F.R.D. at 575.

[15] *Id.* at 577.

[16] *Id.* at 578.

[17] First, the plaintiff "should identify the missing party with sufficient specificity such that the Court can determine that defendant is a real person or entity who could be sued in federal court," *id.* at 578; second, the plaintiff should "identify all previous steps taken to locate the elusive defendant, *id.* at 579; third, plaintiff should show that its claims "could withstand a motion to dismiss," *id.*; and fourth, the plaintiff should specify (and justify) the discovery requests and the entities to which those requests would be addressed, *see id.* at 580.

## C

### *Registrants over whom* in personam *jurisdiction is unavailable*

In addition to the problem of anonymous registrants, the ACPA's drafters also intended to tackle cases in which "a non-U.S. resident cybersquats on a domain name that infringes upon a U.S. trademark."[21] To this end, § 1125(d)(2)(A)(ii)(I) provides that the *in rem* action is also available if the mark owner is unable to obtain *in personam* jurisdiction over the registrant.[22] The problem with this provision is that–as demonstrated below–there exist no cases of foreign cybersquatting as to which § 1125(d)(2)(A)(ii)(I) will be both applicable and constitutional. In order for a court to have territorial jurisdiction in a particular case, there must be a basis for jurisdiction, and the exercise of jurisdiction must be constitutional. A review of the pertinent rules shows that there will always be a basis for *in personam* jurisdiction over ACPA claims against foreign registrants, so long as the exercise of such jurisdiction is constitutional. Thus, § 1125(d)(2)(A)(ii)(I)'s requirement that *in personam* jurisdiction be unavailable will be satisfied only in cases where the exercise of *in personam* jurisdiction would violate due process. In such cases, however, the exercise

---

[18] *See* 15 U.S.C. § 1125(d)(2)(A)(ii)(II).

[19] *See* Mullane v. Central Hanover Bank & Trust Co., 339 U.S. 306, 314 (1950) (requiring "notice reasonably calculated, under all the circumstances, to apprise interested parties of the pendency of the action and afford them an opportunity to present their objections"). Although the ACPA's notice provisions will probably fail to provide actual notice to a registrant who provides false or incomplete contact information to the dealer, or who fails to keep that information current, such a failure should not raise a due process problem. *Cf.* Lehr v. Robertson, 463 U.S. 248, 264 (1983) (approving statutory notice scheme, despite its failure to provide actual notice to appellant, because "the right to receive notice was completely within appellant's control").

[20] 145 CONG. REC. S10513-02 (daily ed. Aug. 5, 1999) (statement of Sen. Hatch). Senator Hatch noted that "some have suggested that dissidents or others who are online incognito for similar legitimate reasons might give false information to protect themselves and have suggested the need to preserve a degree of anonymity on the Internet particularly for this reason." *Id.* The *quasi in rem* provision addresses this concern by "decreas[ing] the need for trademark owners to join the hunt to chase down and root out these dissidents or others seeking anonymity on the Net." *Id.*

[21] H.R. REP. NO. 106-412 (1999).

[22] 15 U.S.C. § 1125(d)(2)(A)(ii)(I).

of *in rem* jurisdiction will be unconstitutional as well.

Bases for jurisdiction

A review of existing rules and doctrines indicates that there will generally be a basis for *in personam* jurisdiction over claims against foreign registrants. Such a basis will be unavailable only where the exercise of *in personam* jurisdiction is unconstitutional–a situation addressed in Part __ below.

For suits in federal court,[23] the basis for jurisdiction is found in Federal Rule of Civil Procedure 4, which authorizes service of process on the defendant. If a foreign registrant has minimum contacts with a particular state in the U.S. and if those contacts meet the requirements of the relevant state long-arm statute, the plaintiff can sue in federal district court in that state and can serve the defendant extraterritorially pursuant to Rule 4(k)(1)(A).[24] If the registrant lacks sufficient minimum contacts to be subject to jurisdiction in any particular state–or if the facts of the case do not fit the relevant state's long-arm statute–then the plaintiff can turn to Rule 4(k)(2).[25] That Rule authorizes service of process on a foreign defendant, in a federal question case, so long as the defendant is not subject to jurisdiction in the courts of any state, and so long as the exercise of jurisdiction "is consistent with the Constitution and laws of the United States."[26] The use of Rule 4(k)(2) is "consistent with" the ACPA because nothing in the ACPA forbids worldwide service of process on an *in personam* defendant. However, the use of Rule 4(k)(2) to authorize *in personam* jurisdiction over ACPA claims against a foreign registrant may violate due process, in which event Rule 4(k)(2) is (by its own terms) inapplicable.

---

[23] It appears that the federal and state courts possess concurrent jurisdiction over *in personam* suits under the ACPA. *Cf.* Aquatherm Indus., Inc. v. Florida Power & Light Co., 84 F.3d 1388, 1394 (11th Cir. 1996) (federal courts do not have exclusive jurisdiction over Lanham Act claims). Plaintiffs will sometimes be better advised to sue in federal court, however, because the jurisdictional reach of a state court will in some instances be less extensive than that of a federal court. *See infra* note __. This advantage may help to explain the apparent dearth of state-court ACPA cases (a Westlaw search of the "allstates" database on August 18, 2001 revealed no cases involving ACPA claims). For purposes of simplicity, this article focuses on ACPA suits brought in federal court.

[24] *See* FED. R. CIV. P. 4(k)(1)(A) (authorizing service of process on defendant "who could be subjected to the jurisdiction of a court of general jurisdiction in the state in which the district court is located").

[25] Neither the ACPA nor the Lanham Act addresses the question of service of process for *in personam* actions. *See, e.g.*, ISI Int'l, Inc. v. Borden Ladner Gervais LLP, 256 F.3d 548, 550 (7th Cir. 2001) (Lanham Act does not authorize worldwide service of process); Quokka Sports, Inc. v. Cup Int'l Ltd., 99 F. Supp.2d 1105, 1110 (N.D. Cal. 1999) (same, with respect to claims under Lanham Act and ACPA). Thus, Rule 4(k)(1)(D)–which permits service of process "when authorized by a statute of the United States"–is inapplicable.

[26] FED. R. CIV. P. 4(k)(2).

In sum, Rule 4 will always provide a basis for *in personam* jurisdiction over ACPA claims against foreign registrants, unless the exercise of such jurisdiction would be unconstitutional. It is to the constitutional analysis, thus, that we now turn.

## 2

## Constitutionality of *in personam* jurisdiction

*In personam* suits against foreign registrants may be constitutional in a number of situations, including cases where the domain name was registered with a U.S.-based dealer. In other instances–as where the registrant uses a foreign-based dealer–due process requirements will often not be met, because the registrant will lack minimum contacts with the United States and the exercise of *in personam* jurisdiction will be unreasonable.

At the outset, a brief discussion of terminology may be helpful. "Registrars" are entities authorized by ICANN to register domain names on behalf of registrants;[27] they function as intermediaries between the individual registrants and the domain name "registry." While there are now multiple registrars (not all of which are based in the U.S.), each TLD has only one registry, which maintains the single authoritative set of records concerning domain names and their registrants.[28] Verisign Global Registry Services, a Virginia-based corporation,[29] operates the registry for the ".com," ".org," and ".net" TLDs,[30] which account for over 85 % of all current domain names.[31] Thus, the relevant registrar may be either a U.S.-based or a foreign corporation, but the pertinent registry for most current domain names is controlled by a U.S.-based corporation. To help distinguish between the two types of entities, we will generally refer to the registrar as the "dealer" and the registry as the "administrator."

To meet the requirements of due process, the defendant must possess minimum contacts with the United States[32] and the exercise of jurisdiction must not be unreasonable. The minimum contacts requirement–which is designed to prevent the assertion of jurisdiction over a defendant having no significant "contacts, ties, or relations" with the forum[33]–is satisfied when a defendant "purposefully directed" its actions at the forum and the litigation arises out of or relates to those acts.[34] Several factors support the argument that a registrant who uses a U.S.-based *dealer* to acquire a domain name thereby creates minimum contacts with the U.S. Although the registrant may communicate with the dealer solely over the Internet, the Court has held that minimum contacts may be found even when the defendant never physically entered the forum.[35] It seems likely that most registrants will be aware of the nationality of the dealer they use. The dealer's website will usually provide reasonable notice that the dealer is a U.S.-based corporation, and may even reveal the specific location of the dealer's physical headquarters. In instances where a reasonable person would infer from the dealer's website that the dealer is U.S.-based, registrants who contract with that dealer to register a domain name can be seen as purposefully directing their activities to the United States.[36]

---

[27] *See* Fleetboston Financial Corp. v. Fleetbostonfinancial.com, 38 F.Supp.2d 121, 123 (2001).

[28] *See id.*

[29] Verisign's website indicates that it is headquartered in Virginia. *See* http://www.verisign-grs.com/aboutus/contact.html (visited August 17, 2001).

[30] *See* Smith v. Network Solutions, Inc., 135 F.Supp.2d 1159, 1163.

[31] *See* http://www.verisign-grs.com/dns/ (visited August 17, 2001) (asserting that .com, .org and .net domain names "comprise over 85% of registered domain names worldwide").

[32] Although the Supreme Court has reserved judgment on the question, *see* Omni Capital International, Ltd. v. Rudolf Wolff & Co., 484 U.S. 97, 102 n.5 (1987) (unanimous opinion); Asahi Metal Indus. v. Superior Court, 480 U.S. 102, 113 n.* (1987) (plurality opinion), it appears that, when a foreign defendant is sued under a federal statute authorizing worldwide service of process, the court may aggregate all of the defendant's United States contacts in order to assess whether the assertion of jurisdiction would comport with due process under the Fifth Amendment. *See, e.g.*, Go-Video v. Akai Elec. Co., 885 F.2d 1406, 1416 (9th Cir. 1989).

The allegations by which the plaintiff seeks to meet the ACPA's bad faith element[37] may establish further connections between the registrant and

---

Thus, in cases where the defendant lacks minimum contacts with the state in which the district court sits, but has contacts with other parts of the United States, the contacts can be aggregated to satisfy the minimum contacts analysis under Rule 4(k)(2). *See, e.g., ISI Int'l*, 256 F.3d at 551 (holding that federal court can exercise jurisdiction under Rule 4(k)(2) over defendant who has "ample contacts with the nation as a whole, but whose contacts are so scattered among states that none of them would have jurisdiction"). Likewise, because the ACPA's *quasi in rem* section provides for worldwide service of process, *see* 15 U.S.C. § 1125(d)(2)(A) & (B), this article assumes that a federal district court asserting jurisdiction under that section should assess whether the defendant possesses minimum contacts with the United States as a whole, rather than with the state in which the district court sits.

[33] International Shoe Co. v. Washington, 326 U.S. 310, 319 (1945).

[34] Keeton v. Hustler Magazine, Inc., 465 U.S. 770, 774 (1984); *see also* Helicopteros Nacionales de Colombia, S.A. v. Hall, 466 U.S. 408, 414 (1984). "General jurisdiction"–which exists when a defendant's contacts with the forum are sufficiently extensive to support jurisdiction over claims unrelated to the contacts–will usually not be available in ACPA cases involving foreign cybersquatters.

[35] *See Burger King*, 471 U.S. at 476.

[36] Although a defendant's contract with a forum resident will not always suffice to establish minimum contacts, such contacts may be shown by the circumstances of the contract. *See Burger King*, 471 U.S. at 478-79; McGee v. International Life Ins. Co., 355 U.S. 220, 223 (1957); Compuserve, Inc. v. Patterson, 89 F.3d 1257, 1264-65 (6th Cir. 1996). In ACPA cases, the defendant will have entered into a contract with a U.S. dealer; the registration will have affected U.S. commerce; and the defendant may have shown an intent to damage the U.S. business of the holder of a mark protected under U.S. law.

[37] The statute makes "bad faith" an element of *in personam* ACPA claims. *See* 15 U.S.C. § 1125(d)(1). The ACPA's *in rem* provision does not explicitly mention bad faith. However, it authorizes a suit *in rem* if (1) the domain name violates the plaintiff's trademark rights and (2) the plaintiff is unable to obtain *in personam* jurisdiction over, or is unable to locate, "a person who would have been a defendant in a civil action under" the ACPA's *in personam* provisions. 15 U.S.C. § 1125(d)(2)(A). A number of courts have concluded that this reference to the *in personam* provisions incorporates the bad faith element into the *in rem* claim as well. *See Broadbridge*

the United States. In assessing whether the plaintiff has properly alleged that the registrant acted with a "bad faith intent to profit from [the plaintiff's] mark,"[38] the ACPA advises the court to consider several factors, including "the [registrant's] intent to divert consumers from the mark owner's online location to a site accessible under the domain name,"[39] and "the [registrant's] offer to transfer, sell, or otherwise assign the domain name to the mark owner or any third party for financial gain."[40] Where a registrant takes such actions against a U.S.-based mark owner,[41] the registrant can be seen as intending to cause an effect within the United States, thus creating contacts for jurisdictional purposes.[42] Moreover, a registrant's choice of a ".com" domain name (rather than a domain name based on a country-code TLD) may sometimes suggest an intent to target U.S. markets.[43]

It should be noted, however, that in five of the six cases to address the question to date[44] the court has held that a foreign defendant's registration of a domain name with a U.S. dealer does not create minimum contacts sufficient to confer *in personam* jurisdiction on a federal court in the district where the dealer is located. One early decision under the ACPA did indicate (without discussion) that a registrant's action in registering the pertinent domain name with NSI (a Virginia corporation) sufficed "to satisfy due process" for purposes of *in personam* jurisdiction.[45] The five subsequent

---

*Media, L.L.C. v. Hypercd.com*, 106 F. Supp.2d 505, 511 (S.D.N.Y. 2000) (bad faith is an element of ACPA *in rem* claims); *Harrods Limited v. Sixty Internet Domain Names*, 110 F. Supp.2d 420, 425 (E.D. Va. 2000) (following *Broadbridge Media*); *Hartog & Co. AS v. Swix.com*, 136 F. Supp.2d 531, 539 (E.D. Va. 2001) (following *Harrods*).

[38] 15 U.S.C. § 1125(d)(1)(A)(i).

[39] *Id.* § 1125(d)(1)(B)(i)(V).

[40] *Id.* § 1125(d)(1)(B)(i)(VI).

[41] Foreign holders of U.S. trademarks can also sue under the ACPA; but a foreign plaintiff would presumably have to show effects on U.S. commerce in order to state a claim. *Cf.* [Vanity Fair etc.]

[42] *See* Panavision Int'l, L.P. v. Toeppen, 141 F.3d 1316, 1322 (9th Cir. 1998) (finding minimum contacts with California, under Calder v. Jones, 465 U.S. 783 (1984), because defendant knew that scheme of registering plaintiff's trademarks as domain names would have "the effect of injuring [plaintiff] in California where [plaintiff] has its principal place of business and where the movie and television industry is centered").

[43] *See, e.g.*, Quokka Sports, Inc. v. Cup Int'l Ltd., 99 F.Supp.2d 1105, 1111-12 (N.D. Cal. 1999) (finding that New Zealand defendants targeted the United States when, instead of choosing a ".nz" domain name, they registered a ".com" domain name with a U.S.-based dealer; defendants "admitted that they sought out a specific domain name to target the 'lucrative American market'").

[44] The issue has been addressed in six published opinions, by three district judges and one magistrate judge, in the Eastern District of Virginia. *See infra* notes __ - __. One of the cases, Heathmount A.E. Corp. v. Technodome.com, 106 F. Supp.2d 860 (E.D. Va. 2000), is currently on appeal to the United States Court of Appeals for the Fourth Circuit.

[45] *See* Lucent Technologies, Inc. v. Lucentsucks.com, 95 F. Supp.2d 528, 531 n.5 (E.D. Va. 2000).

– 10 –

decisions, however, have held to the contrary.[46] The courts finding a due process violation have reasoned that "the utility of a domain name depends in part on the registrar's meeting its obligations, and in part on the operation of the [domain name system], only a small portion of which falls within the domain name registrar's control."[47] Moreover, the typical domain name registration transaction is brief, is conducted over the Internet, involves no negotiation of terms, and does not require the dealer to perform "substantial services" in its home state.[48] These analyses generally appear to assume that the minimum contacts analysis should look to the registrant's contacts with a particular state, rather than aggregating all of the registrant's contacts with the United States. As noted above, that assumption is erroneous;[49] and the courts' preoccupation with assessing the registrant's contacts with the state of Virginia, rather than with the United States as a whole, may have altered some factors in the analysis. Thus, for instance, while it may be true that the registrant of a ".com" domain name would be unaware that NSI is located in Virginia,[50] it is far less plausible that such a registrant would be unaware that it was dealing with a U.S. dealer. On the whole, however, it does not appear that nationwide aggregation of contacts would have altered the conclusion of these courts that registration with a U.S. dealer does not suffice to create minimum contacts.

Whether or not the use of a U.S.-based *dealer* creates minimum contacts, it seems clear that the involvement of a U.S.-based *administrator*, without more, should not create the requisite contacts.[51] Registrants typically have no direct interaction with the administrator. Thus, a French registrant might use a dealer based in France to register a ".com" domain name, unaware that the administrator that will record the domain name is located in the United States. Unless other factors indicate that the registrant aimed its acts at the United States, such a registrant lacks sufficient contacts with the forum to justify the exercise of jurisdiction over the domain name.

In any event, even if minimum contacts exist, a defendant can nonetheless secure dismissal if it can show that the exercise of jurisdiction would nonetheless be unreasonable,[52] based on a five-factor test that considers the burden on the defendant, the forum's interest in hearing the dispute, the plaintiff's interest in obtaining relief, the international judicial system's interest in the efficient resolution of controversies, and the shared interests of various nations in furthering substantive social policies.[53] The burden on a foreign defendant of litigating an ACPA claim in the United States admittedly will be considerable.[54] In addition, the policies of other nations with respect to the regulation of trademarks, and domain names in particular, may differ substantially from those of the United States; and a United States court's adjudication of an ACPA claim may contravene such policies.[55] Balanced against the burden on the defendant and the effect on other countries' trademark policies, however, are the United States' interest in adjudicating the dispute and the plaintiff's interest in obtaining effective relief.[56] The ACPA grew out of congressional concern that U.S. businesses lacked recourse against cybersquatters, including foreign cybersquatters. Its remedies are presumably available only to holders of a mark protected under United States law; and though neither the plaintiff nor the defendant need be a U.S. citizen, the Act applies only in cases where the bad-faith registration has a significant effect on U.S. commerce.[57] So long as the plaintiff brings suit in a district within the state where the dealer is located, the reasonableness analysis may on balance favor the exercise of jurisdiction.

---

[46] *See* America Online, Inc. v. Huang, 106 F. Supp.2d 848, 857 (E.D. Va. 2000) (holding that domain name registration agreements did not "create a sufficient relationship between [the registrant] and Virginia to satisfy due process"); Heathmount A.E. Corp. v. Technodome.com, 106 F.Supp.2d 860, 865 (E.D. Va. 2000) (following *America Online*); Banco Inverlat, S.A. v. www.inverlat.com, 112 F. Supp.2d 521, 522 n.1 (E.D. Va. 2000) (following *Heathmount*); Hartog & Co. AS v. swix.com, 136 F. Supp.2d 531, 536 n.5 (E.D. Va. 2001) (following *Heathmount*); *Cable News Network L.P., L.L.L.P. v. Cnnnews.com*, 2001 WL 1111193, at *3 n.16 (E.D. Va. Sept. 18, 2001) (citing *America Online* and *Heathmount*).

[47] *America Online*, 106 F. Supp.2d at 853.

[48] *America Online*, 106 F. Supp.2d at 855 n.21; *Heathmount*, 106 F. Supp.2d at 866-67.

[49] *See supra* note __.

[50] *See Heathmount*, 106 F. Supp.2d at 866 & n.7.

[51] The fact that no published opinion to date addresses the latter question suggests that potential plaintiffs agree with this assessment.

---

[52] *See* Burger King Corp. v. Rudzewicz, 471 U.S. 462, 487 (1985) (indicating that defendant has burden of demonstrating unreasonableness).

[53] When the Court enunciated these five factors in *World-Wide Volkswagen*, 444 U.S. at 292 (dictum), and applied them in *Burger King*, 471 U.S. at 476-77, 482-84, it was evaluating state courts' assertions of jurisdiction over defendants located outside the forum state but within the U.S. Accordingly, the Court described the last two factors in terms applicable to interstate, rather than international, disputes: "the interstate judicial system's interest in obtaining the most efficient resolution of controversies" and "the shared interest of the several States in furthering fundamental substantive social policies." *World-Wide Volkswagen*, 444 U.S. at 292. The Court has since noted that the application of these two factors to the assertion of jurisdiction over a foreign defendant "calls for a court to consider the procedural and substantive policies of other nations whose interests are affected by the assertion of jurisdiction." Asahi Metal Indus. Co., Ltd. v. Superior Court of California, 480 U.S. 102, 115 (1987) (holding California state court's assertion of jurisdiction over Japanese defendant unreasonable under the circumstances).

[54] *See* Asahi Metal Indus. Co. v. Superior Court, 480 U.S. 102, 114 (1987) ("The unique burdens placed upon one who must defend oneself in a foreign legal system should have significant weight in assessing the reasonableness of stretching the long arm of personal jurisdiction over national borders.").

[55] On the other hand, *Asahi*'s treatment of the reasonableness factors also suggests that the social policies of other nations may weigh more heavily in the defendant's favor when a state court asserts jurisdiction than when a federal court asserts jurisdiction under a federal statute. The *Asahi* Court noted that a state court's assertion of jurisdiction over an alien defendant must be assessed in the light of the federal government's interest in guiding foreign relations. *See Asahi*, 480 U.S. at 115. Where Congress has enacted legislation authorizing suit against a foreign cybersquatter, a federal court's assertion of jurisdiction may be less open to question because the concern of state interference with federal foreign policy does not arise.

[56] *See id.* ("When minimum contacts have been established, often the interests of the plaintiff and the forum in the exercise of jurisdiction will justify even the serious burdens placed on the alien defendant."). It could also be argued that the fifth factor–the international interest in efficient dispute resolution–favors the plaintiff, because an ACPA suit provides a means to determine the rights of each party in the relevant domain name.

Constitutionality of *in rem* jurisdiction

As we have seen, the exercise of *in personam* jurisdiction over ACPA claims against foreign registrants will sometimes be constitutional, but in other cases the exercise of jurisdiction will violate due process. In the latter instances, § 1125(d)(2)(A)(ii)(I) purports to make *in rem* jurisdiction available. Contrary to the apparent expectations of the ACPA's drafters, however, due process requires that there be "minimum contacts" between the registrant and the forum, no matter whether the ACPA claims are denominated *in personam* or *in rem*. In cases where the assertion of *in personam* jurisdiction would violate due process, the assertion of *in rem* jurisdiction would likewise be unconstitutional.

The drafters of the ACPA apparently assumed that a foreign registrant who violated the ACPA's prohibition on bad-faith registration of a domain name[58] would lack minimum contacts with the forum sufficient to satisfy the due process requirements for the exercise of personal jurisdiction,[59] but that the *in rem* provision would help to fill this gap. The ACPA's drafters predicted that *in rem* suits would not offend due process, "since the property and only the property is the subject of the jurisdiction, not other substantive personal rights of any individual defendant."[60] In keeping with this view, a number of courts applying the ACPA have accepted the notion that *in rem* jurisdiction is available despite the absence of minimum contacts for *in personam* purposes.[61] Such a conclusion, however, contravenes the Supreme Court's statement in *Shaffer v. Heitner*[62] that all assertions of jurisdiction[63]–whether *in personam*, *in rem* or *quasi in rem*–must meet the minimum contacts requirements developed in *International Shoe*[64] and its progeny.

---

[57] Cases under other provisions of the Lanham Act indicate that one of the major factors in determining the Lanham Act's reach is whether the defendant's alleged conduct had a significant effect on U.S. commerce. *See, e.g.*, Buti v. Perosa, 139 F.3d 98, 104 n.2, 105 (2d Cir. 1998) (affirming dismissal of foreign defendant's Lanham Act counterclaim because defendant failed to use its mark in commerce in the United States and the mark was not a famous mark); Nintendo of America, Inc. v. Aeropower Co., Ltd., 34 F.3d 246, 249 n.5 (4th Cir. 1994) (Lanham Act reaches extraterritorial conduct "which has a significant effect on United States Commerce").

[58] The ACPA also prohibits bad-faith trafficking in or use of domain names, *see* 15 U.S.C. § 1125(d)(1)(A), but for purposes of simplicity the discussion in the text focuses on the case of bad-faith registration. The jurisdictional issues raised in suits alleging trafficking or use would be similar to those in cases of bad-faith registration; if anything, the case for jurisdiction might be stronger in trafficking or use cases, to the extent that such activities provided additional contacts between the defendant and the United States.

[59] *See* H.R. REP. NO. 106-412 (1999) (stating that "personal jurisdiction cannot be established over the domain name registrant" when the registrant is not a U.S. resident).

[60] H.R. REP. NO. 106-412.

*Shaffer* involved attachment jurisdiction–also known as *quasi in rem* Type 2 jurisdiction–but the *Shaffer* Court made clear that the principles it set forth also apply to *in rem* and *quasi in rem* Type 1 jurisdiction. In *Shaffer*, the Delaware state courts took jurisdiction of a shareholders' derivative suit against officers and directors of a Delaware corporation,[65] based on the attachment, pursuant to a Delaware statute, of corporate stock and options owned by the individual defendants.[66] (Under a Delaware statute, the stock of a Delaware corporation was deemed to be located within the state for purposes of attachment.[67]) The plaintiff alleged that the individual defendants had breached their duties to the corporation by causing the corporation and a subsidiary to engage in activities in Oregon that led to a civil damages award and a large criminal contempt fine.[68] The Delaware courts denied the defendants' jurisdictional challenge, reasoning that *quasi in rem* jurisdiction, which traditionally was based on attachment of property within the jurisdiction, did not require that the defendants have contacts with the forum.[69] The Supreme Court, however, reversed, rejecting both the jurisdictional conclusion and its premise. Noting that under *International Shoe Co. v. Washington*,[70] "the relationship among the defendant, the forum, and the litigation" had become "the central concern of the inquiry into personal jurisdiction,"[71] the Court proceeded to consider whether the

---

[61] *See Heathmount*, 106 F. Supp.2d at 867-68 (finding insufficient contacts for the exercise of *in personam* jurisdiction, but allowing *quasi in rem* claim to proceed); *Banco Inverlat*, 112 F. Supp.2d at 522 n.1 (same); *Hartog*, 136 F. Supp.2d at 536 & n.5 (same); *Cable News Network*, 2001 WL at *4 (same).

[62] 433 U.S. 186 (1977).

[63] The statement in *Shaffer* pertained to "assertions of state-court jurisdiction," 433 U.S. at 212, but the Court's reasoning is equally applicable to the exercise of jurisdiction by federal courts. *See infra* text accompanying notes _ – _.

[64] International Shoe Co. v. Washington, 326 U.S. 310 (1945).

[65] The suit named as defendants Greyhound Corp., Greyhound's wholly owned subsidiary Greyhound Lines, Inc., and 28 current or former officers or directors of one or both entities. *Shaffer*, 433 U.S. at 189-90.

[66] *Shaffer*, 433 U.S. at 190-94.

[67] *Id.* at 192 n.9.

[68] *Id.* at 190.

[69] *Id.* at 196.

[70] 326 U.S. 310 (1945).

[71] *Shaffer*, 433 U.S. at 204.

*Shoe* standard "should be held to govern actions *in rem* as well as *in personam*."[72] Because "judicial jurisdiction over a thing" (the traditional conception of *in rem* jurisdiction) is merely "a customary elliptical way of referring to jurisdiction over the interests of persons in a thing,"[73] the Court concluded that "in order to justify an exercise of jurisdiction *in rem*, the basis for jurisdiction must be sufficient to justify" *in personam* jurisdiction–i.e., it must meet "the minimum-contacts standard elucidated in *International Shoe*."[74] The Court made clear that this standard applied to all assertions of *in rem* jurisdiction, not just to the type of *quasi in rem* jurisdiction that was at issue in *Shaffer* itself: the Court stated flatly that "all assertions of state-court jurisdiction must be evaluated according to the standards set forth in *International Shoe* and its progeny," and it added that "[t]o the extent that prior decisions are inconsistent with this standard, they are overruled."[75]

Consistent with its stated intention to set a standard for application to all *in rem* cases, the Court took pains to assess the likely effect of its new approach on different types of *in rem* jurisdiction. Under *Shoe*, although the mere presence of property within the forum will not in itself justify jurisdiction, it is not irrelevant, for it can help to provide the requisite minimum contacts between the defendant and the forum. As the Court put it, "when claims to the property itself are the source of the underlying controversy between the plaintiff and the defendant, it would be unusual for the State where the property is located not to have jurisdiction," since "the defendant's claim to property located in the State would normally indicate that he expected to benefit from the State's protection of his interest."[76] Moreover, in such cases the forum will often have "strong interests in assuring the marketability of property within its borders and in providing a procedure for peaceful resolution of disputes about the possession of that property," and relevant evidence and witnesses will often be found within the forum[77]–factors which would support the argument that the exercise of jurisdiction would be reasonable. Accordingly, the Court concluded that its extension of the *Shoe* standard to all assertions of state-court jurisdiction appeared unlikely to affect jurisdiction over most *in rem* actions other than those in which the property attached was unrelated to the claim.[78]

The Court recognized, however, that in *quasi in rem* Type 2 cases such as *Shaffer* itself, the imposition of the *Shoe* standard *would* "result in significant change," because the defendant's ownership of property within the forum would be unrelated to the plaintiff's cause of action.[79] Where the cause of action does not relate to or arise out of the defendant's contacts with the forum, those contacts will not meet the *Shoe* standard unless they are continuous and systematic–a test that will not be met by the mere ownership of property within the forum.[80] In *Shaffer*, the Court concluded that neither the defendants' ownership of stock in the Delaware corporation nor their positions as officers or directors of that corporation provided the requisite minimum contacts for purposes of the shareholders' derivative suit; accordingly, it held that the Delaware courts' assertion of jurisdiction violated due process.[81]

---

[72] *Id.* at 206.

[73] *Id.* at 207 (quoting Restatement (Second) of Conflict of Laws § 56, Introductory Note (1971)).

[74] *Id.*

[75] *Id.* at 212 & n.39.

[76] *Id.* at 207-08.

[77] *Id.* at 208.

[78] *Id.*

Although it might at first seem that the ACPA's *in rem* provisions satisfy the minimum contacts analysis sketched out in *Shaffer*, such an argument cannot withstand scrutiny. The argument would be that a plaintiff can bring an ACPA *in rem* suit only when the domain name was registered by a United States dealer or administrator; that in such instances the ACPA deems the domain name to be property located within the United States; and that the plaintiff's claim thus arises directly out of a claim to the registrant's property located within the forum. However, as the *Shaffer* Court noted, even in cases where the plaintiff's cause of action arises out of or relates to the defendant's claim of ownership of the pertinent property, the presence of that property within the forum will not always support the inference of contacts between the defendant and the forum. (The Court suggested that such an inference might be unfounded, for example, in cases where a chattel was brought into the forum without the owner's consent or where the plaintiff's fraud induced the owner to send the chattel into the forum.[82])

The *Shaffer* Court's caveat foreshadows the issues raised by the ACPA. The ACPA's *in rem* provision authorizes suits against domain names registered with a U.S.-based dealer *or* a U.S.-based administrator (such suits are to be brought "in the judicial district in which the domain name registrar, domain name registry, or other domain name authority that registered or assigned the domain name is located"[83]). Assuming that Congress has the authority to designate domain names as a form of property that can be subjected to attachment for purposes of *in rem* jurisdiction, and assuming further that Congress has the authority to provide that such domain names are located within the United States whenever the dealer or administrator involved in registering the domain name is located within the U.S., the resulting "presence" of the domain name within the United States does not, without more, provide sufficient minimum contacts between the registrant and the United States.[84] If, for example, the registrant registered the domain name with a foreign dealer and had no idea that the domain name would be administered by a United-States-based registry, the "presence" of the domain name within the United States would not indicate the existence of minimum contacts between the registrant and the United States. In sum, the "presence" of the domain name within the United States adds nothing to the minimum contacts analysis that would apply to a suit *in personam*; either the registrant has sufficient minimum contacts to satisfy due process, or else the *Shoe* standard will bar the exercise of any type of jurisdiction, either *in rem* or *in personam*.[85]

---

[79] *Id.* at 208.

[80] Continuous and systematic contacts with the forum have been held sufficient for the exercise of jurisdiction over an unrelated claim against a corporation. *See Perkins v. Benguet Consol. Mining Co.*, 342 U.S. 437, 447-48 (1952).

[81] *See Shaffer*, 433 U.S. at 213-17.

[82] *See Shaffer*, 433 U.S. at 208 n.25 (citing Restatement (Second) of Conflict of Laws § 60, cmts. c & d).

[83] 15 U.S.C. § 1125(d)(2)(A).

The result of the application of *Shaffer* to the ACPA *in rem* provision is that the provision is of use only in cases involving anonymous registrants; in cases where, instead, the registrant is known but cannot be subjected to an *in personam* ACPA claim, the registrant's lack of minimum contacts with the United States will similarly bar the assertion of *in rem* jurisdiction. Two district courts in the Eastern District of Virginia have resisted this conclusion, arguing that *Shaffer* does not require the application of minimum contacts analysis to ACPA *in rem* actions.[86] For example, the court in *Cable News Network L.P., L.L.L.P. v. Cnnews.com*[87] ("*CNN*") held that "in an ACPA *in rem* action, it is not necessary that the allegedly infringing registrant have minimum contacts with the forum."[88] The arguments advanced to support this assertion fall into three general categories: that *Shaffer*'s requirement of minimum contacts is *dictum* as applied to the *in rem* cause of action created by the ACPA and can thus be disregarded;[89] that Justice Scalia's plurality opinion in *Burnham v. Superior Court*[90] somehow overruled *Shaffer*'s requirement of minimum contacts for *in rem* actions; and that other authorities provide some basis for a refusal to apply the minimum contacts test to ACPA *in rem* claims. Each of these arguments, however, fails.

---

[84] *Cf.* Fleetboston Financial Corp. v. Fleetbostonfinancial.com, 138 F. Supp.2d 121, 134 (D. Mass. 2001).

[85] Shaffer also forecloses the argument that the limited nature of the remedies available through an in rem suit loosens the requirements of due process, *see, e.g.*, *America Online*, 105 F. Supp.2d at 858 n.32 (holding that registration of a domain name did not create minimum contacts sufficient for in personam jurisdiction, and distinguishing a quasi in rem ACPA case on the ground that "the registrant's contact with [the registrar] satisfied due process" in light of "the limited relief available under the in rem proceeding, namely forfeiture of the domain name in question"). While the *Shaffer* Court recognized that "the potential liability of a defendant in an in rem action is limited by the value of the property," it found this fact irrelevant to the due process analysis because the fairness of subjecting the defendant to jurisdiction "does not depend on the size of the claim being litigated." *Shaffer*, 433 U.S. at 207 n.23.

[86] *See Cable News Network L.P., L.L.L.P. v. Cnnews.com*, 2001 WL 1111193, at *4 (E.D.Va. Sept. 18, 2001); *Caesars World, Inc. v. Caesars-palace.com*, 112 F.Supp.2d 502, 504 (E.D.Va. 2000). It appears that the drafters of the ACPA relied on this contention as well. *See* H.R. REP. NO. 106-464 (1999).

[87] 2001 WL 1111193 (E.D. Va. Sept. 18, 2001).

[88] *Id.* at *4.

[89] The *CNN* court reasoned that "*Shaffer*'s language regarding true *in*

As demonstrated above, the *Shaffer* Court clearly intended to extend the *International Shoe* framework to all cases of *in rem* jurisdiction; and whether that extension was dictum or holding, it should be applied to the *in rem* provisions of the ACPA.[91] Notably, the *Shaffer* Court itself characterized as a "holding" its conclusion "that any assertion of state-court jurisdiction must satisfy the *International Shoe* standard."[92] Moreover, *Shaffer*'s core principle–that jurisdiction over a thing is merely another way of describing jurisdiction over the interests of persons in that thing–applies with equal force to all *in rem* cases, whether or not the cause of action is related to the property that forms the basis for jurisdiction. Indeed, recognizing this, the majority in *Shaffer* analyzed the probable effects of the holding on cases in which "claims to the property itself are the source of the underlying controversy."[93] Even if *Shaffer*'s statement is dictum as it applies to *in rem* and *quasi in rem* Type 1 cases,[94] it is carefully considered dictum. When the Supreme Court articulates a general principle of constitutional doctrine, and especially when the Court takes pains–as it did in *Shaffer*–to assess the implications of that principle for contexts other than that of the case at hand, lower courts should be slow to brush the principle aside as mere "dictum."[95]

---

rem and *quasi in rem* I matters was unnecessary to the holding and is therefore non-binding dicta." *Cable News Network*, 2001 WL at *4.

[90] 495 U.S. 604 (1990).

[91] *See* Fleetboston Financial Corp. v. Fleetbostonfinancial.com, 138 F. Supp.2d 121, 134 (D. Mass. 2001) ("The logic of *Shaffer*'s limitations would appear to extend to actions in which the existence of the property in the state cannot fairly be said to represent meaningful contacts between the forum state, the defendant, and the litigation. While this will generally be type II quasi in rem actions, it will not be so exclusively.").

[92] *Shaffer*, 433 U.S. at 208 (stating that "[i]t appears . . . that jurisdiction over many types of actions which now are or might be brought *in rem* would not be affected by a holding that any assertion of state-court jurisdiction must satisfy the *International Shoe* standard.").

[93] *Id.* at 207. Although the Court then proceeded to focus on the application of the minimum contacts test to the assertion of *quasi in rem* Type 2 jurisdiction, that focus arose not only from the fact that *Shaffer* itself involved *quasi in rem* Type 2 jurisdiction, but also from the Court's judgment that "acceptance of the *International Shoe* test would most affect this class of cases." *Id.* at 209.

[94] It is not self-evident that the statements concerning these types of *in rem* jurisdiction should be viewed as dictum. *See* Michael C. Dorf, *Dicta and Article III*, 142 U. PA. L. REV. 1997, 2040 (1994) (arguing that the distinction between holding and dictum should turn on the rationale articulated by the court, rather than simply on "facts and outcomes").

[95] It is a truism that dictum does not constitute binding precedent. As Chief Justice Marshall stated in *Cohens v. Virginia*, "general expressions, in every opinion, are to be taken in connection with the case in which those expressions are used. If they go beyond the case, they may be respected, but ought not to control the

The *CNN* court acknowledged that "there is language in *Shaffer* that could be read to require that *all in rem* cases conform to the same due process constraints as *in personam* cases," but asserted that "the greater weight of (and more persuasive) authority holds that the language of *Shaffer* requires minimum contacts only for *quasi in rem* II-type cases."[96] The authorities referred to, however, are either inapposite or erroneous. For example, the court cites Justice Scalia's opinion in *Burnham v. Superior Court*,[97] a case in which the Court upheld a state court's assertion of *in personam* jurisdiction over a defendant who was personally served with process while physically present in the forum state. In the portion of the *Burnham* opinion cited by the *CNN* court, Justice Scalia (joined by Chief Justice Rehnquist and Justice Kennedy) argued that the *Shoe* minimum contacts analysis need not be applied to the *Burnham* defendant, and distinguished *Shaffer* on the basis that it involved an "absent defendant," rather than one who is physically present within the forum state at the time of service of process: "The logic of *Shaffer*'s holding–which places all suits against absent defendants on the same constitutional footing, regardless of whether a separate Latin label is attached to one particular basis of contact–does not compel the conclusion that physically present defendants must be treated identically to absent ones."[98] Notably, Justice Scalia's description of *Shaffer*'s "logic" actually supports *S h a f f e r*'s application to "all suits against absent defendants"–including ACPA *in rem* actions. It is true that Justice Scalia also argued that the result in *Burnham* should turn on the historical pedigree of "tag" jurisdiction; as Justice Scalia noted, this reliance on tradition

---

judgment in a subsequent suit when the very point is presented for decision." *Cohens v. Virginia*, 19 U.S. 264, 399 (1821). Nonetheless, although the precise standard varies from circuit to circuit, the lower federal courts customarily accord substantial weight to Supreme Court dictum, particularly when that dictum was carefully considered by the Court. *See, e.g., City of Timber Lake v. Cheyenne River Sioux Tribe*, 10 F.3d 554, 557 (8th Cir. 1994); *Guidry v. Sheet Metal Wkrs. Int'l Assoc., Local No. 9*, 10 F.3d 700, 706 n.3 (10th Cir. 1993) ("If [Supreme Court] dicta had clearly resolved the issue in this appeal we would be bound by that decision."); *McCoy v. Massachusetts Institute of Technology*, 950 F.2d 13, 19 (1st Cir. 1991) ("[F]ederal appellate courts are bound by the Supreme Court's considered dicta almost as firmly as by the Court's outright holdings, particularly when . . . a dictum is of recent vintage and not enfeebled by any subsequent statement."); *Nichol v. Pullman Standard, Inc.*, 889 F.2d 115, 120 n.8 (7th Cir. 1989) (Court of Appeals "should respect considered Supreme Court dicta"); *United States v. Underwood*, 717 F.2d 482, 486 (9th Cir. 1983) (in banc) ("The Supreme Court cannot limit its constitutional adjudication to the narrow facts before it in a particular case. In the decision of individual cases the Court must and regularly does establish guidelines to govern a variety of situations related to that presented in the present case. The system could not function if lower courts were free to disregard such guidelines whenever they did not precisely match the facts of the case in which they were announced."); *Jordon v. Gilligan*, 500 F.2d 701, 707 (6th Cir. 1974) ("Even the Court's dicta is [sic] of persuasive precedential value."); *Fouts v. Maryland Casualty Co.*, 30 F.2d 357, 359 (4th Cir. 1929) ("[D]icta of the United States Supreme Court should be very persuasive."); *cf. United States v. Bell*, 524 F.2d 202, 206 (2d Cir. 1975) (stating that "considered or 'judicial dictum' where the Court . . . is providing a construction of a statute to guide the future conduct of inferior courts," though not binding, "must be given considerable weight").

[96] *Cable News Network*, 2001 WL at *4.

[97] 495 U.S. 604 (1990).

[98] *Burnham*, 495 U.S. at 621.

contradicts the Court's approach in *Shaffer*, which applied minimum contacts analysis to *in rem* jurisdiction despite its "ancient form[]."[99] Justice Scalia's reasoning in *Burnham*, however, would not validate the use of the ACPA's *in rem* procedure; as Justice Scalia acknowledged, "[f]or new procedures, hitherto unknown," the due process inquiry is guided by *International Shoe*.[100] Although *in rem* jurisdiction has a long historical pedigree, the same cannot be said of the application of *in rem* jurisdiction to Internet domain names. Rather, because the ACPA's attempt to use domain names and Internet contacts as a basis for jurisdiction is an indisputably modern construct, Justice Scalia's appeal to tradition in *Burnham* would provide no support for the constitutionality of the ACPA's *in rem* provisions. (In any event, no part of Justice Scalia's *Burnham* opinion commanded a majority of the Justices.)

Of the other authorities cited by the *CNN* court, the only sources that directly support *CNN*'s holding[101] are one district court case and one law

---

[99] *Burnham*, 495 U.S. at 622.

[100] *Id.* at 622.

[101] The *CNN* court cites four other authorities which are inapposite.

In *Amoco Overseas Oil Co. v. Compagnie Nationale Algerienne de Navigation*, 605 F.2d 648 (2d Cir. 1979), the court upheld the assertion of *quasi in rem* Type 1 jurisdiction because the attached funds were payments that the plaintiff had made to the defendant with respect to the contract at issue in the case, and the payments were in the relevant New York bank account pursuant to the contract, *see id.* at 655. The court also noted that there was no indication the defendant would be amenable to suit anywhere else in the world, and thus the doctrine of jurisdiction by necessity supported the exercise of jurisdiction. *See id.* Finally, the court reasoned that "jurisdiction by attachment of property should be accorded special deference in the admiralty context." *Id.* (Issues concerning jurisdiction by necessity are discussed below, *see infra* text accompanying notes __ - __.)

In *Schreiber v. Allis-Chalmers Corp.*, 611 F.2d 790 (10th Cir. 1980)–a case involving the assertion of *in personam* jurisdiction–the Tenth Circuit held that *Shaffer* did not undermine the holding of *Perkins v. Benguet Consolidated Mining Co.*, 342 U.S. 437 (1952). In *Perkins*, the Court had held that a corporation that carries on continuous and systematic activities within a state can be sued in the courts of that state on a claim unrelated to the corporation's in-state activities. *See Perkins*, 342 U.S. at 447-48. Far from being at odds with the *Shoe* requirement of minimum contacts, *Perkins* articulates the nature of that requirement in cases of general (as opposed to specific) jurisdiction. *Shaffer*'s statement that the *Shoe* analysis governs all state-court assertions of jurisdiction does not undermine *Perkins*, which specifies how the *Shoe* analysis applies in cases of general jurisdiction. Thus, the court's statement in *Schreiber* that "*Shaffer* is distinguishable," *Schreiber*, 611 F.2d at 793, is both correct and utterly irrelevant to the question presented by the ACPA.

In *John N. John, Jr., Inc. v. Brahma Petroleum Corp.*, 699 F. Supp.2d 1220 (W.D. La. 1988), the court found *Shaffer* "inapposable" because "the property attached is the very subject of the cause of action," *id.* at 1222. Since the property in question was tangible and was shipped into the jurisdiction by the plaintiff on behalf of the defendant, *see id.* at 1220, the property's presence within the forum might well be seen to provide contacts between the defendant and the forum. Moreover, the court found that "sufficient review article. In *Caesars World, Inc. v. Caesars-palace.com*,[102] the court rejected the argument that the minimum contacts requirement applies to the assertion of *in rem* jurisdiction under the ACPA, reasoning that *Shaffer* is limited to cases in which the cause of action "is unrelated to the property which is located in the forum state."[103] This argument, which is merely a restatement of the "*Shaffer* as dictum" argument, fails for the reasons discussed above. The law review article, likewise, relies mainly on the contention that Justice Scalia's *Burnham* opinion weakens the force of *Shaffer*[104]–an assertion which, as noted, is unpersuasive.

There remain two possible arguments not yet advanced by the courts that have rejected *Shaffer*'s application to ACPA cases: first, that *Shaffer* by its terms applies only to assertions of jurisdiction by state, not federal courts, and second, that, even if *Shaffer* ordinarily would require minimum contacts for ACPA *in rem* suits, such suits should be allowed to proceed under the doctrine of jurisdiction by necessity. Neither argument, however, is likely to succeed. Admittedly, the limitations imposed by the due process clause of the Fifth Amendment on the exercise of territorial jurisdiction by federal courts differ in some respects from the limitations imposed on state courts through the due process clause of the Fourteenth Amendment. For instance, as noted above,[105] it appears likely that the Fifth Amendment due process analysis can in appropriate cases look to an absent defendant's contacts with the United States as a whole, rather than just to the defendant's contacts with the state in which the federal court sits. However, in other respects the doctrines of federal court territorial jurisdiction draw heavily upon the due process analysis developed under the Fourteenth Amendment; and though the *Shaffer* Court referred only to state court jurisdiction, the logic of the opinion supports a similar analysis with respect to federal court jurisdiction

---

contacts" existed because the defendant had purposefully availed itself of the privilege of conducting activities within the state, by contracting to do business within the state. *Id.* at 1222.

Finally, the *CNN* court cites a law review article that proposes "a new constitutional test for personal jurisdiction." *See* Walter W. Heiser, *A "Minimum Interest" Approach to Personal Jurisdiction*, 35 WAKE FOREST L. REV. 915, 915 (2000).

[102] 112 F. Supp.2d 502 (E.D. Va. 2000).

[103] *Id.* at 504. Despite its assertion that "it is unnecessary for minimum contacts to meet personal jurisdiction standards," the *Caesars World* court proceeded to address the question of minimum contacts, and concluded that "the fact of domain name registration with Network Solutions, Inc., in Virginia supplies" such contacts. *Id.* at 504.

[104] *See* Thomas R. Lee, *In Rem Jurisdiction in Cyberspace*, 75 WASH. L. REV. 97, 137-40 (2000). Lee also seems to suggest that *Shaffer* itself does not require minimum contacts analysis in *in rem* or *quasi in rem* Type 1 proceedings, *see id.* at 141-42, and asserts that in any event the registration of a domain name with NSI in Herndon, Virginia provides the requisite minimum contacts for an ACPA suit, *see id.* at 142-43.

[105] *See supra* note __.

as well. Accordingly, *Shaffer*'s requirement that the defendant have minimum contacts with the forum should apply equally to federal court assertions of *in rem* jurisdiction under the ACPA (although the minimum contacts analysis would focus on the registrant's contacts with the United States as a whole). Nor does the doctrine of jurisdiction by necessity validate the ACPA's *in rem* provision. That doctrine has been argued to support a court's exercise of jurisdiction over defendants, despite the defendants' lack of minimum contacts with the forum, if no other court would have territorial jurisdiction over the defendants. For example, the Court in *Shaffer* noted, but did not consider, "whether the presence of a defendant's property in a State is a sufficient basis for jurisdiction when no other forum is available to the plaintiff."[106] The Court has subsequently indicated, however, that such a theory is unavailable in cases where the defendant is subject to suit in a foreign court.[107] The courts of the country where a domain name registrant is located presumably will have territorial jurisdiction over claims against that registrant.[108] Moreover, domain name registrants are subject to non-judicial proceedings under ICANN's Uniform Domain-Name Dispute-Resolution Policy, and those proceedings provide the same remedies to a successful claimant as would the ACPA's *in rem* provisions. Accordingly, the doctrine of jurisdiction by necessity seems unlikely to validate those provisions.

In sum, contacts between a registrant and the United States will suffice for *in rem* jurisdiction only if the contacts are extensive enough to support *in personam* jurisdiction as well. If the requisite minimum contacts exist, *in personam* jurisdiction will (as discussed above) be available, and thus § 1125(d)(2)(A)(ii)(I) will not apply. As a result, the only cases in which § 1125(d)(2)(A)(ii)(I) provides a basis for *in rem* jurisdiction are those in which the exercise of jurisdiction would violate due process. Once the courts recognize this constitutional problem, the ACPA's *in rem* provision will be of no use to a mark owner seeking to sue a foreign domain name registrant who lacks minimum contacts with the United States. Indeed, the provision already is of little use, as evidenced by the fact that most mark owners choose to proceed under the ICANN Uniform Domain-Name Dispute-Resolution Policy instead of suing in federal court under the ACPA. Thus, the significance of the *in rem* provision lies mainly in Congress's aggressive assertion of jurisdiction over domain name disputes involving foreign registrants, and in Congress's attempt to ground the exercise of jurisdiction in the purported "presence" of the domain name within the United States. The assertion that the domain name is located within the U.S. in turn depends on the location of the dealer or administrator within the U.S. Similarly, as we discuss below, the strength of the U.S.'s claim to prescriptive jurisdiction over international domain name disputes depends largely on the present geographical location of the registry in charge of administering the key top-level domains. But while the *in rem* provision's use of dealer and administrator locations is of conceptual rather than practical interest (since the *in rem* provision is unconstitutional anyway), the significance of geography in the prescriptive jurisdiction analysis has real-world consequences.

---

[106] *Shaffer*, 433 U.S. at 211 n.37.

[107] In *Helicopteros Nacionales de Colombia, S.A. v. Hall*, 466 U.S. 408 (1984), the Court rejected the plaintiffs' jurisdiction by necessity argument on the ground that the plaintiffs "failed to carry their burden of showing that all three defendants could not be sued together in a single forum . . . . for example, . . . in either Colombia or Peru," *id.* at 419 n.13.

[108] Admittedly, the fact that a foreign court would have territorial jurisdiction over the registrant does not mean that the foreign court would also have subject matter jurisdiction over an ACPA claim against that registrant, or that foreign law would provide any similar remedy. However, these questions are academic, because the availability of the UDRP procedures (discussed in the text) should remove any argument that jurisdiction by necessity validates the ACPA's *in rem* provisions.

## III

### Naming Jurisdiction:
### Realspace Sovereigns & Domain Names

For the balance of this paper, we turn from the adjudicative to the legislative, and from the constitutional to the strategic. In this section we consider the implications of the present far-reaching Congressional approach to legislative jurisdiction over domain name disputes, noting that such a U.S.-centric view may not in the long term be in the interests of either the global electronic commerce community as a whole, or the U.S. participants in this new economic structure in particular.

### *A*

### *Distributed Hierarchy: the Control of Domain Names*

As presently constituted,[109] the domain name system[110] is nothing more (and nothing less) than a distributed hierarchical database – a simple list of names and their corresponding IP addresses.[111] That is, no single

---

[109] circa Fall 2001.

[110] We'll assume here that readers have a working understanding of the domain name system – at least enough to it use in the ordinary course of web-browsing, for example. Brief, non-technical introductions to the domain name system can be found in a variety of places, both online and off. *See, e.g.,* Margaret Jane Radin & R. Polk Wagner, *The Myth of Private Ordering: Rediscovering Legal Realism in Cyberspace*, 73 CHI.-KENT L. REV. 1295, 1303 (1999); Diane Cabell, *Name Conflicts*, LEARNING CYBERLAW IN CYBERSPACE (1999), http://www.cyberspacelaw.org/cabell/index.html.

[111] Internet communications are "packet-based", meaning that transmissions are separated into small data units, wrapped in addressing (and other) information, and sent across the Internet. In order to reach their destination, packets must be

computer contains the entire database; the computers that do contain the database (called "DNS servers") are located in myriad locations (both physically and logically) worldwide. Yet the hierarchical nature of the system means that some parts of the database are more important than others. To illustrate this point, we'll consider the system as having three distinct levels: the "root" level, the "TLD" level, and the "user" level, as shown in Figure 1.

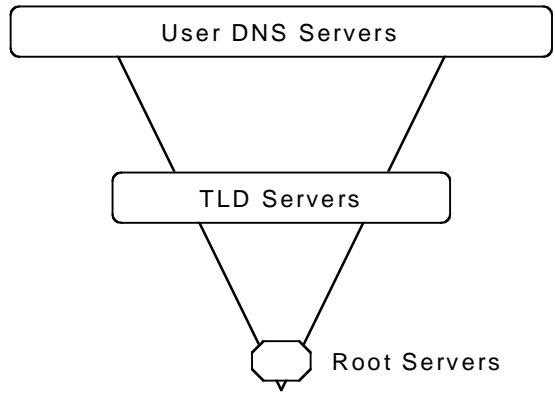

**Figure 1: Domain Name Hierarchy**

The "user" DNS servers are those that serve individual groups of users/machines – subscribers to a particular ISP, for example, or those within a corporate or university network. That is, the "user" DNS servers for the University of Pennsylvania contain the addressing information for the machines relating to "upenn.edu" (Penn's assigned domain name), as well as serve DNS queries sent by these machines.[112] The "TLD Servers" hold the addressing information for the user DNS servers about an entire top-level-domain – here, the ".edu" TLD server would contain (among others) the addressing information for the upenn.edu DNS server.[113] And the "root servers" hold addressing information relating to the TLD servers; for

---

addressed with the appropriate "IP address" – a unique number corresponding to each machine connected to the Internet. When an Internet user requests a page using a domain name (say, for example, *www.yahoo.com*, the domain name system provides the correct IP address.

[112] For example, the IP address for the machine with the name *law.upenn.edu* is 130.91.144.200. This information is maintained by the Penn DNS servers, the primary one of which is located at 128.91.254.4.

[113] This information is easily found by conducting a "whois" query on the appropriate database. A web-based user interface to one such database can be found at http://www.netsol.com/cgi-bin/whois/whois. The *whois* record for upenn.edu contains the following DNS data:

*Domain servers in listed order:*

| | |
|---|---|
| *noc3.dccs.upenn.edu* | 128.91.254.4 |
| *noc2.dccs.upenn.edu* | 128.91.254.1 |
| *dns1.udel.edu* | 128.175.13.16 |
| *dns2.udel.edu* | 128.175.13.17 |

example, the location of the ".edu" TLD server.[114] Because DNS requests for out-of-network resources (say a user in the Penn network requests the address corresponding to *www.yahoo.com*) will in theory[115] require requests to each level of the DNS system described above, it becomes apparent that the hierarchy of the system determines the amount of requests to which the various servers must respond. That is, the .edu TLD server will (absent caching) be involved in all out-of-network requests involving the .edu domain, and the root server will be involved in essentially all DNS requests.

This distributed hierarchy, then, has unquestionable regulatory significance: involvement in DNS requests means an ability to exert power, at least over the next higher (in our Figure 1) DNS level. Indeed, the regulatory significance of the DNS hierarchy is what the Internet Corporation for Assigned Names and Numbers (ICANN), the present administrator of the public root server system, uses to ensure universal adoption of the Uniform Dispute Resolution Policy (UDRP) – because ICANN controls the root servers, it determines the status of the TLD servers. Further, by requiring any applicant desiring to administer a TLD to agree to require its domain name registrants to either agree to the provisions of the UDRP, or gain approval for any alterations from the UDRP, ICANN can essentially exert full regulatory control over the domain name system.[116]

---

[114] As of October 2001, the TLD servers for .edu were:

*a.root-servers.net.*

*h.root-servers.net.*

*c.root-servers.net.*

*g.root-servers.net.*

*f.root-servers.net.*

*b.root-servers.net.*

*i.root-servers.net.*

*e.root-servers.net.*

*d.root-servers.net.*

[115] As a practical matter, local DNS servers often "cache" recent requests.

[116] *See* ICANN, New TLD Application Process Overview § (August 3 2000) (http://www.icann.org/tlds/application-process-03aug00.htm)("For unsponsored TLDs, ICANN will have policy-formulation responsibility for the new TLD and the policies will initially be generally defined as the existing policies for .com, .net, and .org . . . . ."). Of course ICANN can (and does) delegate some policymaking authority to TLD operators, most prominently in the case of "sponsored" TLD – where the sponsoring organization is delegated some range of policymaking authority. *See* ICANN, New TLD Program (2001) (http://www.icann.org/tlds/) ("Generally speaking, an "unsponsored" TLD operates under policies established by the global Internet community directly through the ICANN process, while a "sponsored" TLD is a specialized TLD that has a sponsor representing the narrower community that is most affected by the TLD. The sponsor thus carries out delegated policy-formulation responsibilities over many matters concerning the TLD.). Nonetheless, the control (provided by administration of the root server system) is there.

B

*Mapping Control: Geography and the Domain Name System*

Thus far we have described what might be called "logical" control over the domain name system: the technological ability to exert power by virtue of administration of the important components of the DNS system.[117] And yet while this alone raises a number of interesting questions,[118] we want to raise a slightly different set of issues: those surrounding the realspace sovereigns' approach to regulation of the domain name system.

Here, we want to point out that the domain name system unquestionably exists in realspace. It is merely a collection of computer hardware and software (and people who administer and maintain them); each of these constituent components is "real" in any sense of the word, and can be found at various geographic points throughout the world. As such, it is possible to map the domain name system onto realspace, where concepts like borders and sovereignty are crucially important (at least from a regulatory perspective).

For example, we noted above that the primary root server, named *a.root-servers.net*, is administered by ICANN. A simple web query reveals that machine is physically located within the United States, in the city of Herndon, Virginia.[119] That machine also serves as the .edu TLD server.[120] The primary user DNS server for *upenn.edu*, named *noc3.dccs.upenn.edu*, is located in Philadelphia, Pennsylvania – as expected.[121]

It follows, then, that if the distributed hierarchy of the domain name system has logical regulatory significance, and this hierarchy is mapped onto realspace, the geography of the domain name system has what might be called "territorial" regulatory significance. It is to the implications of this territorial regulatory significance that we now turn.

---

[117] This general view – the correlation between control over technology and policymaking – is a common theme in recent "cyberlaw" scholarship. For an excellent overview, see Lessig, CODE AND OTHER LAWS OF CYBERSPACE (1999).

[118] In particular we note the many issues that surround the establishment of ICANN as the authoritative body for the present public root server system. Jonathan Weinberg, *ICANN and the Problem of Legitimacy*, 50 Duke L.J. 187 (2000); Milton Mueller, *ICANN and Internet Governance: Sorting through the debris of self-regulation*. INFO 1, 6, 477-500, December 1999. See also the excellent materials relating to ICANN collected by ICANN Watch, a watchdog organization, at http://www.icannwatch.org/ [more cites]

[119] NetGeo Query, found at http://netgeo.caida.org/perl/netgeo.cgi. The results also noted the approximate latitude and longitude as 38.98 / -77.39.

[120] *See supra* note __. The other major TLD servers – those for the .com, .org, and .net TLDs – are also located near Herndon, Virginia – Dulles, to be exact. Each of these TLDs has as its primary DNS server the machine designated *a.gtld-servers.net*. *See* IANA, .com Top-Level Domain, Root-Zone Whois Information, http://www.iana.org/root-whois/com.htm, *See* IANA, .net Top-Level Domain, Root-Zone Whois Information, http://www.iana.org/root-whois/net.htm, *See* IANA, .org Top-Level Domain, Root-Zone Whois Information, http://www.iana.org/root-whois/org.htm.

[121] Latitude 39.96, longitude –75.20.

*C*

*The Significance of Territorial Regulation*

This territorial significance – that the geographic location of certain important components of the domain name system will have implications for the regulation of the system – plays out in at least two ways. First, as an international legal matter, the geographic location of the domain names system may be widely recognized as supporting a strong claim to prescriptive (or legislative) jurisdiction over various disputes that arise relating to the system. Second, even if a prescriptive jurisdictional claim is not widely recognized according to accepted principles, the geographic location of the domain name system may nonetheless determine whether (and to what extent) realspace sovereigns can regulate the system. Yet this point -- that the geography of the domain name system will to a large degree determine the scope of regulatory authority available to realspace sovereigns -- must be considered in the context of the essentially arbitrary nature of the present geographic facts of the system. That is, while the logical regulatory significance of the hierarchy of the domain name system is inherent in the technology, the geography is not. Put simply, the root servers could move, taking with them the territorial regulatory significance. Furthermore, a realspace sovereign[122] set on increasing the present regulatory authority available to itself – either recognized or *de facto* – could itself alter the geographic facts, by creating additional root servers and requiring their use. Recognition of this requires, we think, a widespread reconsideration of the aims of present regulation of the domain name system by realspace sovereigns, in particular the United States' approach embodied in the ACPA. We take up this final point in Section __ below, while first discussing the dual implications of territoriality and the domain name system.

Recognized Authority: Prescriptive Jurisdiction & Domain Names

We turn first (and briefly) to the significance of prescriptive jurisdictional claims.[123]

---

[122] Or set of sovereigns, of course. For example, a group of sovereigns, perhaps organized regionally or culturally, might establish and mandate its own root server system, with shared or delegated policy authority.

[123] As will become quickly apparent to the reader, this subsection is not intended as a comprehensive description of the issues surrounding prescriptive jurisdiction. Rather, our goal here is to provide context to support our argument that the geography of the domain system will have realspace regulatory implications.

### a
### The Principles of Prescriptive Jurisdiction

As a general matter, international law recognizes at least three, and as many as five bases of prescriptive jurisdiction.[124] The most widely accepted principles relevant to our discussion[125] are those of:

territoriality;[126] and,

nationality.[127]

"Territoriality" refers to the notion that a sovereign state has a claim to prescribe conduct within its physical territory,[128] and the "status of persons, or interests in things, present within its territory."[129] This basis for jurisdiction includes the related "effects principle," which extends the basic claim to include "conduct outside [the state's] territory that has or is intended to have substantial effect within its territory." [130] Although some commentators have suggested that the territorial principle should be the complete basis for analyzing prescriptive jurisdiction[131], the modern view

---

[124] We use the terms "legislative jurisdiction" and "prescriptive jurisdiction" interchangeably here, given the context in which we write: the analysis of the scope of Congressional authority to legislate on the matter of domain names. See, e.g., RESTATEMENT OF THE LAW, THIRD, FOREIGN RELATIONS LAW OF THE UNITED STATES, § 401(a) [hereafter, RESTATEMENT].

[125] As we note below in notes __ and accompanying text, there are several other recognized principles of prescriptive jurisdiction; none of which seem likely to be relevant in the context of the dispute over domain names.

[126] RESTATEMENT § 402(1), Barry Carter & Philip Trimble, INTERNATIONAL LAW 728, 729-733 (2nd ed., 19__) [hereafter, Carter & Trimble ]

[127] RESTATEMENT § 402(2), Carter & Trimble, 728,

[128] "It is an essential attribute of the sovereignty of this realm, as of all sovereign independent States, that it should possess jurisdiction over al persons and things within its territorial limits and in all causes civil and criminal arising within these limits." J. Starke, INTRODUCTION TO INTERNATIONAL LAW 194 (9th ed. 1984) (quoting Lord Macmillan).

Mark Janis suggests that the treaties concerning the Peace of Westphalia of 1648 established "the principles of the territorial sovereignty and the jurisdiction of states have been two of the most fundamental principles of international law." Janis, INTRODUCTION TO INTERNATIONAL LAW (2nd ed., 19 __).

[129] RESTATEMENT 402(1)(b).

[130] RESTATEMENT 402(1)(c). See also United States v. Aluminum Company of America, 148 F.2d 416, 443 (2nd Cir. 1945) ("[A]ny state may impose liabilities, even upon persons not within its allegiance, for conduct outside its borders that has consequences within its borders which the state reprehends; and these liabilities other states will ordinarily recognize."); Starke, 195; Carter & Trimble, 738 ("[I]n the past few decades, U.S. law enforcement agencies and U.S. Courts have increasingly applied U.S. law extraterritoriality, that is, to persons acting abroad, when their acts have a substantial effects on the United States.");

[131] "[T]he character of an act as lawful or unlawful must be determined wholly by the law of the country where the act is done. . . . For another jurisdiction, if it should happen to lay hold of the actor, to treat him according to its own notions rather than those of the place where he did the acts, not only would be unjust, but would be an interference with the authority of another sovereign, contrary to the comity of

recognizes that the territorial principle is both under- and over-inclusive: it both fails to account for cases in which states have a legitimate claim to prescriptive jurisdiction, and must yield in certain circumstances to the extraterritorial principles that have developed to address the under-inclusiveness (including, indeed, the "effects principle").[132]

The principle of "nationality" is also rooted in traditional notions of state sovereignty, though in this case the "sovereignty" refers to the state's citizens or subjects, rather than the state's physical territory.[133] As typically stated, nationality refers to the right to prescribe "the activities, interests, status, or relations of its nationals outside as well as within its territory."[134]

Note that three other commonly-invoked bases for prescriptive jurisdiction exist: (a) the protective principle; (b) passive personality principle ; and, (c) the universality principle. Under the protective principle, a state may prescribe conduct occurring outside its territory and not performed by its nationals, if such conduct threatens the security of the state or certain other special classes of state functions.[135] While this basis for jurisdiction is well-established, the challenge in delimiting the scope of state interests that may be used to invoke the principle – as well as the self-referential aspects of any such definition – often leads to controversy over its application.[136] Here, it appears unlikely that it would support the assertion of prescriptive jurisdiction over domain names.

"Passive personality" refers to the claim that a state may exercise jurisdiction whenever one of its citizens is harmed, irrespective of the

---

nations, which the other state concerned justly might resent. . . ." American Banana Co. v. United Fruit Co., 213 U.S. 347, 356-357 (1909) (Holmes, J.).

[132] This recognition, however, is not without controversy. Carter & Trimble note that application of U.S. antitrust and securities law to acts committed abroad has sparked resistance from foreign courts, and even the passage of "blocking statutes" in various states intended to curtail the extraterritorial reach of U.S. laws. Carter & Trimble, 738. The Restatement addresses this dispute by expressly limiting prescriptive jurisdiction according to a standard of reasonableness. RESTATEMENT § 403. See also notes __ and accompanying text.

[133] See, e.g., Blackmer v. United States, 284 U.S. 421 (1932) ("Nor can it be doubted that the United States possesses the power inherent in sovereignty to require the return to this country of a citizen, resident elsewhere, whenever the public interest requires it, and to penalize him in the case of refusal.");

[134] RESTATEMENT 402(2).

[135] RESTATEMENT 402(3). The special classes of state functions are usually taken to support jurisdiction over subject such as espionage, counterfeiting of the state's seal or currency, falsification of official documents, as well as perjury before consular officials, and conspiracy to violate the immigration or customs laws. See Restatement 402, comment f.

[136] To take one example, the protective principle has been used to uphold the right of the United States to prosecute foreign nationals on foreign vessels on the high seas for possession of narcotics. United States v. Romero-Galue, 757 F.2d 1147 (11th Cir. 1985). Starke notes that while the grounds for the protective principle presume that the offenses subject to its application are (1) of "utmost gravity" to the state and (2) may escape punishment altogether, absent the assertion of jurisdiction, the fact that each state has the latitude to determine what conduct implicates important state functions means that the application of the protective principle can be "quite arbitrary".

location of the dispute or the identity of the harmed.[137] This claim is not widely recognized beyond circumstances involving international terrorism and other serious crimes, however,[138] and we'll not discuss it further here. Similarly, the "universality principle" recognizes that some crimes – usually limited to those offenses "recognized by the community of nations as of universal concern, such as piracy, slave trade, attacks on or hijacking of aircraft, genocide, war crimes, and perhaps certain acts of terrorism" – may be prescribed anywhere.[139] Again, this principle is not relevant to our discussion here.

### b
### Limitations to Prescription

As should be apparent, even confining the discussion to the two most widely-accepted and relevant bases for prescriptive jurisdiction – territoriality and nationality – does little to reduce the potential for competing claims among states. Accordingly, the claims outlined above must be understood to be limited in certain respects. To this end, The Restatement (Third) of Foreign Affairs sets forth the general United States approach, which is in essence a two-part test: first, where a competing claim for prescriptive jurisdiction might exist, the exercise of jurisdiction must be "reasonable"; second, where two states have "not unreasonable" claims, jurisdiction should be given to the state with a "clearly greater" interest in prescription.[140] Under the restatement, both the reasonableness of the claim of jurisdiction and the weighing of the states' interests are measured by a non-exclusive set of eight factors, generally involving the extent and quality of the connection between the location, the parties, and the states involved – as well as the character of the dispute itself.[141] As applied by the United States, these limitations are most clearly reflected in canons established for the construction of statutes that may involve a contested claim of prescriptive jurisdiction.[142] For example, there is the "longstanding principle of American law 'that legislation of Congress, unless a contrary intent appears, is meant to apply only within the territorial jurisdiction of the United States.'" [143] Even

---

[137] RESTATEMENT 402, comment g.; Starke, 224; Carter & Trimble, 782.

[138] RESTATEMENT 402, comment g ("The principle has not been generally accepted for ordinary torts or crimes, but it is increasingly accepted as applied to terrorist and other organized attacks on a state's nationals by reason of their nationality, or to assassination of a state's diplomatic representatives or other officials."); United States v. Columba-Collela, 604 F.3d 356 (5th Cir. 1979).

[139] RESTATEMENT 404; Carter & Trimble, 788-89; Janis, 329-30. See also United States v. Smith, 18 U.S. 153, 161 (1820) (jurisdiction to prescribe piracy), Filartiga v. Peña-Irala, 630 F.2d 876, 890 (2nc. Cir. 1980) (jurisdiction to prescribe torture).

[140] RESTATEMENT § 403(1), (3).

[141] RESTATEMENT § 403(2).

[142] E.g., Hartford Fire Insurance v California, 509 U.S. 764, 813-17 (1993) (Scalia, J. dissenting)

[143] EEOC v. Arabian American Oil Co., 499 U.S. 244, 248 (1991) (quoting Foley Bros., Inc. v. Filardo, 336 U.S. 281, 285 (1949))( finding the presumption not overcome

more significantly, Chief Justice Marshall established the canon that "[a]n act of Congress ought never to be construed to violate the law of nations if any other possible construction remains."[144] Taken together, this analysis seeks to limit the circumstances where the application of U.S. prescriptive jurisdiction will run into conflict with the claims of another state, and explicitly draws in the examination of reasonableness and the respective strengths of the claims involved.[145] Accordingly, courts in the United States have limited the application of U.S. laws where they determined that to do otherwise would undermine the "interacting interests of the United States and of foreign countries,"[146] and otherwise advocated caution when presented with contesting claims of prescriptive jurisdiction.[147]

A further, yet related, limitation on the claims underlying prescriptive jurisdiction noted above is the principle of "comity". Comity refers to the mutual respect sovereigns offer one another by limiting application of their laws in certain cases – most commonly where another nation has a more significant claim to jurisdiction.[148] In many ways, the application of principles of comity parallels the "reasonableness" inquiry suggested by the Restatement – at root, they both involve an analysis of the strength of the sovereign interests. What comity adds to this analysis is a

---

in the case of Title VII of the Civil Rights Act of 1964). See also Matushita Elec. Indus. Co. v. Zenith Radio Corp., 475 U.S. 574, 582, n. 6 (1986) (approving the extraterritorial application of the Sherman Act), Continental Ore Co. v. Union Carbide & Carbon Corp., 370 U.S. 690, 704 (1962) (same); United States v. Aluminum Co. of America, 148 F.2d 416 (2nd Cir. 1945) (same).

[144] Murray v. Schooner Charming Betsy, 6 U.S. 64 (1804).

[145] See RESTATEMENT § 403.

[146] Romero v. Int'l Terminal Operating Co., 358 U.S. 354, 383 (1959). See also Lauritzen v. Larsen, 345 U.S. 571 (1953), McCulloch v. Sociedad Nacional de Marineros de Honduras, 372 U.S. 10, 21-22 (1963).

[147] United States v. Aluminum Co. of America, 148 F.2d 416, 443 (2nd Cir. 1945) ("[W]e are not to read general words, such as those in [the Sherman] Act, without regard to the limitations customarily observed by nations upon the exercise of their powers; limitations which generally correspond to those fixed by the 'Conflict of Laws.'").

[148] "'Comity' . . . is neither a matter of absolute obligation, on the one hand, nor of mere courtesy and goodwill, upon the other. But it is the recognition which one nation allows within its territory to the legislative, executive or judicial acts of another nation, having due regard both to international duty and convenience, and to the rights of its own citizens or of other persons who are under the protections of its laws." Hilton v. Guyot, 159 U.S. 113, 163-64 (1985).

Comity refers to two types of limitations. First, the limitations that are self-imposed by the legislature in adopting a statute. Justice Scalia has termed this sort of comity "prescriptive comity," and notes that courts will assume that this type of comity has been exercised when evaluating the scope of statutes. See Hartford Fire, at 817. See also Timberlane Lumber Co. v. Bank of America, N. T. & S. A., 549 F.2d 597, 608-615 (9th Cir. 1976); Mannington Mills, Inc. v. Congoleum Corp., 595 F.2d 1287, 1294-1298 (3rd Cir. 1979); Montreal Trading Ltd. v. Amax Inc., 661 F.2d 864, 869-871 (10th Cir. 1981). The second sort of comity is judicial in nature, a recognition by a court that the decision in the case is better made elsewhere, or that the decision must be fashioned in a way that considers foreign interests. Hartford Fire, 817; Carter & Trimble, 738. See also J. Story, COMMENTARIES ON THE CONFLICT OF LAWS § 38 (1834) (distinguishing between the "comity of the courts" and the "comity of nations," and defining the latter as "the true foundation and extent of the obligation of the laws of one nation within the territories of another") suggestion of flexibility, a sort of subjective consideration of how the determination of prescriptive jurisdiction – irrespective of the more objective criteria of "reasonableness" and "clearly greater" interests – fits into the wider context of international legal principles.[149]

<div style="text-align:center">

c
### Prescribing Domain Names

</div>

Having set forth the bases and limitations of the assertion of prescriptive jurisdiction (especially in the U.S. context), we turn now to how these bases interact with the geography of the domain name system. Because we are especially interested in the United States' present regulatory approach (*i.e.*, the ACPA, especially the *in rem* provisions[150]), and because the current geographic facts of the domain name system are remarkably US-based, this discussion will necessarily be heavily US-focused. (Note that in the particular context in which Congress has presently acted – legislating for domain name disputes as they relate to trademarks[151] – there will be one or more registered U.S. Trademarks involved;[152] for purposes of our argument, however, we set aside the international trademark issues.[153])

---

[149] Carter & Trimble, 738.Janis, 332-33.

[150] 15 U.S.C. § 1125(d) (2000). The self-styled "in rem" proceedings are found in section 1125(d)(2).

[151] See 15 U.S.C. § 1125(d).

[152] See 15 U.S.C. § 1125(d) (allowing claims based on infringement of a "mark"). Federal Trademarks are registered by application to the United States Patent and Trademark Office, pursuant to 15 U.S.C. §§ 1051 – 1072.

[153] It is an understatement to note that the application of trademark law to the online context presents difficult international law issues. [ . . . ] As Graeme Dinwoodie has noted, see Graeme Dinwoodie, Private International Aspects of the Protection of Trademarks (2001)(WIPO Doc. No. WIPO/PIL/01/4), the continued expansive interpretation of national trademarks in the online context threatens to "reduce" trademark rights to "their most destructive form" – the mutual ability to block (or at least interfere with) the online use of marks recognized in other countries. Accordingly, the World Intellectual Property Organization (WIPO) is expected to approve in fall 2001 a resolution calling for a more flexible recognition of "use" of trademarks on the Internet, one that would provide protections from liability for legitimate users of marks who disclaimed the intent to conduct commerce in a particular country. See Standing Committee on the Law of Trademarks, Industrial Designs and Geographical Indications, Protection of Marks, and Other Industrial Property Rights in Signs, on the Internet (2001) (WIPO Doc. No. SCT/6/2).

We set aside the issues specific to international application of US trademark law for several reasons. Most generally, we consider the truly notable question here to be Congress' approach to the regulation of domain names, especially the self-styled *in rem* provisions of the ACPA; in this view, the effects on the international regulation of trademarks are parasitic on a particular view of domain names as what one of us has described as "a species of mutant trademark." Margaret Jane Radin & R. Polk Wagner, *The Myth of Private Ordering: Rediscovering Legal Realism in Cyberspace*, 73 CHI.-KENT L. REV. 1295, 1303 (1999). Here, looking beyond the mutant trademark aspects of domain names allows, we hope, for a more focused consideration of the challenges facing sovereigns in cyberspace.

Our analysis suggests that the present form of the domain names system offers the United States a strong claim of prescriptive jurisdiction along several dimensions of the analysis. Maintaining our focus on the bases of nationality and territoriality, we note the following important factors involved in this analysis:

*Citizenship*. The citizenship of the parties to the dispute obviously invokes the nationality principle, supporting the prescriptive claim for any corresponding sovereign.

*TLD Server Location*. Because almost all domain name disputes will arise in the context of what are known as second-level domain names (the "upenn" in the name "upenn.edu" is the second-level domain name), the most direct authority over and responsibility for these domain names will rest with the operation of the relevant TLD Server, thereby supporting the prescriptive claim of the sovereign in which that server is located.[154]

*Root Server Location*. As noted above, control over the root server allows at least some level of control over the entire system, though at times this support will be indirect.[155] As such, the geographic location of the root server will support – though to a lesser degree than the location of the TLD server – the assertion of prescriptive jurisdiction, based on the principle of territoriality.

Focusing first on the United States, we note again that the primary root server (*a.root-servers.net*) is located within the United States,[156] we set forth our results as follows:

---

But there are other, more practical, reasons to set aside the international trademark questions as well. First, to the extent that such questions turn on the identity and citizenship of the parties, *see* Vanity Fair Mills, Inc. v. T. Eaton Co. Ltd., 234 F.2d 633 (2nd Cir. 1956), we consider that in our analysis. Second, the consideration of whether there are conflicting rights to use the mark, see Vanity Mills, is, in our view, distinct from the question of whether there are (or should be) legitimate competing claims to use the domain name. See Dinwoodie, etc. Third, whether there are "substantial effects" in the United States resulting from the use of a trademark as part of a domain name is (a) similar in nature to the "effects" analysis we discuss at note __ below; and (b) likely to be a disputed ground upon which to determine the scope of prescriptive jurisdiction on the Internet. See Playboy Enters., Inc. v. Chuckleberry Pub. Inc., 939 F.Supp 1032 (S.D.N.Y. 1996). Finally, the use of trademark law to support legislation pertaining to domain names is likely to create additional conflicts and controversy is only likely to further buttress our second point made at notes __ and accompanying text below: that the current US approach lends incentives for other states to consider supporting the segmenting of the domain name space.

[154] Note that for simplicity, we're conflating the location of the actual hardware (i.e., the machine) that constitutes the TLD server, and the location of the administrative authority. We assume, for the purposes of our argument, that the hardware will typically be in the same country as the administrators / system operators.

[155] See supra notes __ and accompanying text.

[156] See supra notes __ and accompanying text.

| case | citizenship | TLD Server | Prescriptive Claim |
|------|-------------|------------|---------------------|
| 1 | US | US | US |
| 2 | mixed | US | US |
| 3 | non-US | US | US |
| 4 | US | non-US | US |
| 5 | mixed | non-US | US |
| 6 | non-US | non-US | unclear |

**Table 1: U.S. Prescriptive Jurisdictional Claims**

Cases 1 and 2 are relatively easy: in each, both nationality and territoriality support the U.S. claim; in Case 2, there is a potential competing claim based on nationality, but the clear weight of the interests balances in favor of the U.S. Case 3 is perhaps more controversial. If the non-US citizens are from separate countries, then there are at least three potentially competing claims. Among the three, we suggest that none will have a "clearly greater" interest than the U.S.[157] If the non-US citizens are from the same country, then the relative claim of that country would seem stronger.[158] In Cases 4 and 5, the U.S. claim is supported by nationality, as well as the basis for territoriality provided by the location of the root server.[159] In our view, the location of the TLD server does not provide a "clearly greater" claim in this context, though we admit these cases are close.

Cases 6 has an uncertain result. Of course, the U.S. would have at least some support from the principle of territoriality as a consequence of the location of the root server – though this support would be less, we think, than the support provided by the location of the TLD server. Yet that would be the full extent of the support for the U.S. claim, leading us to conclude that the U.S. claim is likely (though perhaps not certain) to be clearly outweighed by a competing claim, especially if the location of the TLD server and the citizenship of at least one of the parties corresponds.

This exercise leads us to two points. First, even though the U.S. Constitution severely limits the impact of the *in rem* provisions of the ACPA,[160] the U.S. prescriptive jurisdictional claim appears to be quite strong

---

[157] Pursuant to the RESTATEMENT, when multiple countries have competing claims, a "clearly greater" claim must be recognized. See RESTATEMENT § 403(3). *See also* supra notes __ and accompanying text.

[158] *See, e.g., Heathmount v. Technodome.com*, 106 F. Supp.2d 860 (E.D. Va.2000) (ACPA action between Canadian citizens). The court in *Heathmount* did not address this issue.

[159] Note that the analysis of the "effects" of domain name activity is likely to approximate a wash in cases where the DNS system is available worldwide. That is, the effects of a domain name in the United States is likely to be similar to the effects of the same domain name in another country; both states might be able to note the effects in their territory, but it seems unlikely that either effect would be "clearly greater" than the other.

[160] As we analyzed in section __ above, these *in rem* provisions, found in 15 U.S.C. § 1125(d)(2) apply to cases of anonymous registrants and cases involving registrants lacking minimum contacts with the United States. In the latter cases, as we discuss

generally. Second, as far as recognizable claims of prescriptive jurisdiction goes, geography matters: the physical location of the TLD and root servers play a crucial role in evaluating potentially competing claims of prescriptive jurisdiction. Here, the geographic fact that the root servers (at least the primary one and its administration), as well as the most populated TLD servers, are geographically located within the United States, grants the United States a considerable amount of regulatory latitude. And while the U.S. government has, we think, a limited ability to exercise jurisdiction over domain name disputes, other sovereigns may not be so limited.[161] As we note below, this insight has substantial implications.

*De Facto* Regulatory Significance

Even if a particular sovereign's claims for prescriptive jurisdiction are unrecognized under international legal principles, we note that the geography of the domain name system can nonetheless provide substantial regulatory leverage. This de facto regulatory significance flows, like the recognized form, from the distributed hierarchy of the domain name system. To the extent that elements of the domain name system are under a sovereign's potential physical control – typically by being located (geographically) within the sovereign's territory – a sovereign can exercise control. For example, because the administration of the root server offers logical control over the entire domain name system,[162] sovereign control over the root server would then allow de facto control over the domain name system.

For example, consider the case of the United States, under the present geographic and technical facts of the domain name system.[163] In principle, Congress could pass laws (or an agency could issue regulations) directed to the public root server system. Perhaps these laws might specify the standards by which TLDs would get access to the root servers, or even specify the TLDs and their policies themselves. These laws might give preferences to U.S. companies and individuals in domain name disputes, charge taxes on any entity using a domain name, or specify the types of uses that domain names can be put to.[164] By controlling the root server, the U.S.

---

above, the assertion of *in rem* jurisdiction over the domain name will violate due process.

[161] Assuming the same level of geographic connections to the critical features of the domain name system. As we note below, sovereigns can also take steps to alter the geographic facts in their favor.

[162] At least over the domain name system utilizing that root server. As we noted above, this is how ICANN presently exerts nearly complete policy control over the domain name system. See supra notes __ and accompanying text.

[163] See supra notes __ and accompanying text.

[164] Obviously, the U.S. Constitution would provide an important limit to such regulations, especially those that might implicate rights of free expression. See, e.g., U.S. Const. amend. 1. Note, however, that such constraints would not necessarily restrict similar actions by other sovereigns.

government could effectively control the TLD servers – by threatening banishment from the public root server system and the concomitant loss of operation. And by controlling the TLD servers, the U.S. government could exercise de facto control over the entire range of second-level domain names available in the public root server system.

There is no evidence at all, we think, that the United States government is planning any activities of this sort.[165] Nor, of course, are we advocating such a course, as we argue in some detail below. At present, even the most aggressive assertion of jurisdiction we describe here – the *in rem* provisions of the ACPA -- falls well short of the sort of widespread *de facto* control set forth above. And yet the point we noted above remains: geography matters. Put simply, control over the domain name system -- both recognized and de facto – is remarkably correlated with geographic facts. This, we think, has considerable implications for the future of the domain name system.

## IV

## The Crucial Role of Realspace Sovereigns

Thus far, we've analyzed the aspects of logical and territorial control over the domain name system, arguing that in the context of realspace sovereigns, the mapping of the logical control structure onto the physical world results in a regulatory hierarchy of sorts – with those states that have critical components of the domain name system located within their territory having potentially great regulatory leverage over the system as a whole. This argument also leads to two other important observations. First, we note that the geographic facts of the domain name system are both essentially arbitrary and uniquely mutable attributes,[166] thereby providing realspace sovereigns with an important hook to attempt to alter the geographic facts in their favor. Second, and following from the first, we argue that the likely result of greater regulatory activism by various realspace sovereigns will be the segmentation of the domain name system, and the dramatic reduction in utility provided by the system itself.

### *A*

### *Geographic Alteration: The Virtual Land Grab*

---

[165] Cite the White Paper, Memorandum of Understanding.

[166] Note that we do not suggest that that realspace geography ("territoriality") doesn't matter on the Internet generally. It is becoming increasingly obvious that geography is both ascertainable and important online. [cites to NetGeo, Packeteer, Akamai, Yahoo! France ]. Instead, the suggestion is that, given the distributed hierarchical nature of the DNS, the geographic location of the "A" root servers is not especially relevant to the operation of the system. See supra note 30.

Say that Country X decides that the present quasi-U.S. approach to the regulation of the domain names system is not in its interest. Perhaps this country is troubled by the present strong protection of commercial trademarks embedded in both the ICANN-mandated policies and the ACPA.[167] Or perhaps Country X is disagrees with the principles of free expression that have been held to extend to domain name disputes.[168] Or Country X might simply see the domain name system as a potential source of revenue.[169]

Whatever the reason, it is clear that Country X would have substantial ability to at least attempt to alter the geographic facts. Country X could establish a root server system of its own, and mandate that local networks use the "official" root server.[170] In a fairly benign form of this "virtual land grab," the Country X root server might mirror or link to the existing public root server system, and merely offer additional TLDs.[171] Country X would be able to exert regulatory control over the additional TLDs, while the remaining TLDs would be unaffected and remain available to Country X users. A more troubling arrangement would be if Country X limited, reassigned the TLDs available in the new root server, or otherwise created conflicts between its root server and the public server. This second set of circumstances creates the (great, we think) potential for the *segmentation* of the worldwide domain name system, and the dramatic reduction in the value of the system.

B

*Segmenting Domain Names*

As we use it here, the domain name system becomes "segmented" when: (1) the same DNS requests sent by users in different networks yield different results; or (2) some number of TLDs are unavailable to users, depending upon the root server system they use. Both of these circumstances arise when different root server systems are used – though the existence and use of different root servers is a necessary but not a sufficient condition.[172] Segmentation arises when the various root server systems in use are either in conflict, or do not accurately reflect the content of other root servers. For example, the official "Country X Root Server" might send requests for the .com TLD to a different server than the ICANN *a.root-servers.net*, thereby creating a conflict – resulting in, for example, a request for *www.yahoo.com* to yield a different web page in different networks.[173] Or the Country X Root Server might ignore requests for the .com TLD altogether, rendering a large part of the domain name system unusable (for its users). An even more likely case is that any additional TLDs established by Country X may not be recognized by the public root server system, thereby making them unavailable to the Internet user community at large.[174]

---

[167] *See supra* notes __ and accompanying text. *See also* Froomkin, etc.

[168] See, e.g.,

[169] *See, e.g.,* Anna Soderblom, *Island Joins the Dots and Will Net Fortune*, THE TIMES (LONDON), Nov. 6, 2000 (noting that the island nation of Tuvalu expects to net over $50 million from registrations in the .tv TLD space, "or about three times Tuvalu's gross domestic product."). The Tuvalu case is slightly different than what we suggest here, of course, as Tuvalu controls only the .tv TLD, rather than the root server.

[170] Alternatively, the regulations could identify a private "alternative" root server system – perhaps as part of a wider agreement involving policies, taxation, etc. Such alternative root server systems, as we discuss at notes __ and accompanying text below, already exist. See, e.g., New.net, Mission Statement (2001) at http://www.new.net/about_us_mission.tp

[171] This type of arrangement has been described as a "virtual inclusive root". *See* S. Higgs, *Root Zone Definitions*, INTERNET ENGINEERING TASK FORCE (Internet Draft) (May 2001).

We predict that segmentation would result under the following circumstances:

First, a new root server system (formed or mandated by a sovereign's regulatory activity) could create conflicts with the existing public root server system. As noted above, these conflicts could result in unexpected behavior or the inability for certain segments of the global Internet to utilize TLDs publicly available elsewhere.

Second, and perhaps even more importantly, segmentation must be considered a potential response by one or more sovereigns to any others' attempts to exert unwanted regulatory influence over the domain name system. For example, if Country X established its own root server system, and established policies for expanded TLDs that were in conflict with the policies of Country Y, a potential response for Country Y would be to responsively create (or mandate) a root server system that effectively "blocked" the Country X TLDs. Lest this possibility seem farfetched, consider China's recent suggestion that it would take steps to ensure that it controlled the distribution and administration of all Chinese-character domain names – a task that certainly implicates the creation of alternative root server systems (or the threat thereof).[175]

---

[172] As noted in Higg's *Root Zone Definitions*, multiple root server systems presently exist with little, if any, problems for the system. See supra note __. Higgs in particular describes "private" root server systems, which are not publicly available and are intended to serve only a single network, as well as "inclusive" root servers, which expand the TLDs available on the standard root server system. Id.

[173] This would not necessarily be the case, of course. Depending upon the details of Country X's redistribution plan, it seems entirely possible that Yahoo!, Inc. might purchase the rights to yahoo.com in the County X Root Server zone.

[174] Note that we distinguish here between what we call "segmentation" of the domain name system and the present use and availability of "alternative" or "inclusive" root server systems. The major (in terms of numbers of users) alternative root server systems differ from the hypothetical segmented system we outline above because they explicitly include the information provided by the public root server system, and because they are created by and backed with market forces, not sovereign government regulatory action. We recognize that such alternative root server systems may stimulate much-needed innovation and competition in the field of domain name administration, yet note that the increasing popularity of such services is only increasing the chances that true segmentation develops.

175 See, e.g., Rachel Ross, China Demands Jurisdiction Over Domain Names In Chinese, TORONTO STAR, Nov. 13, 2000 ("'We think that as 97.5 per cent of the As should be easily apparent, segmentation of the domain name system would dramatically decrease its value to the Internet user community.[176] As a means of searching and selecting online resources, the domain name system's value is directly related to its scope; its value is at its highest when the system includes all named online resources. Importantly, if the domain name system cannot reliably be considered authoritative of the resources available, its value diminishes remarkably -- if an Internet user does not have confidence that typing *www.yahoo.com* will yield the information she expects, then the DNS request will be of far less value. And while there are good arguments that the domain name system has outlived its usefulness as a uniform means of addressing the Internet searching problem, we think that an unsegmented domain name system remains a significant value to Internet users.[177] This premise – that an unsegmented domain name system is desirable – has substantial policy implications for realspace sovereigns. In the following section, we take up this issue.

### C

*Unsegmenting Policy: Realspace Sovereigns and Domain Names*

---

people using Chinese characters live in the mainland and Taiwan, the U.S. government has no right to authorize any company to manage Chinese domain names with Chinese characters,' said CNNIC director Hu Qiheng . . . . 'A company shouldn't be allowed to provide Chinese domain names registration services in China without the approval of the Chinese government.'"); ., Character debate: CNNIC opposes foreign firms registering Chinese-language domain names, CHINAONLINE, Nov. 3, 2000 (explaining how China is objecting to the registration and use of Chinese-character domain names by foreign entities) (Lexis).; Furious Fight Arises in Registration of Chinese Domain Names, Xinhua General News Service, Nov. 17, 2000 (Lexis) (same); China to strengthen management of Chinese-character domain names, CHINAONLINE, Nov. 15, 2000 (Lexis); China Channel, CNNIC Disagree Over Chinese Character International Domain Names, CHINAONLINE, Oct. 10, 2000 (Lexis) (same). At present, the ICANN-approved registrars are still registering such domain names. *See* Verisign, *Key Points About the VeriSign Global Registry Services Internationalized Domain Name Testbed* (2001) http://www.verisign-grs.com/idn/keypts.html.

[176] This appears to be a nearly universal sentiment. See, e.g., Internet Architecture Board (Network Working Group), IAB Technical Comment on the Unique DNS Root (Request for Comments 2826) (May 2000); ICANN, ICP-3: A Unique, Authoritative Root for the DNS (July 9, 2001) at http://www.icann.org/icp/icp-3.htm; [others]

[177] We think a not insignificant argument can be made that the development of a segmented domain name system would so dramatically devalue the system as to generate alternatives to the searching and selection problem that may prove at least as effective as the present domain name system One obvious example might be an increasing investment in web search engines or directories. Or the development of new technologies, such as the RealNames "keywords" system. See, e.g., RealNames.com. The rise of and eventual replacement by alternatives to the domain name system should not necessarily be viewed as unfortunate. Nonetheless, an unsegmented domain name system appears at present to be high on the list of valuable approaches to the searching problem.

The present United States approach to domain name regulation[178] appears to be focused solely on the potential for harm to a particular form of indigenous commercial interests – namely, trademarks. Little legislative attention has be paid to the significant value in the unsegmented domain name system -- or more particularly, the significant diminishment in value should the system become segmented. This omission becomes all the more serious when one considers that the aggressive assertion of jurisdiction in the ACPA[179] may well satisfy the conditions under which the domain name system becomes segmented: the encouragement of responsive actions on the part of other sovereigns.[180] This is, we think, a perilous path – one that risks harming the growth and development of the global Internet, and correspondingly portends harm to United States interests.

In this section, we briefly discuss the justifications and contours of a more encompassing approach to the regulation of domain names by realspace sovereigns – in particular the United States. Such an approach, we argue, would result in greater deference to and support of the growth of non-territorial regulatory devices for this uniquely global asset.

Encompassing Interests

Given that the growth of ecommerce development thus far has largely redounded to the benefit of the United States' commercial interests,[181] the continued development of the Internet should be viewed as beneficial; concomitantly, obstacles to such growth should be avoided.[182] We've established above that an unsegmented domain name system is significantly more valuable to the global internet community than one that is partially or fully segmented. It follows, then, that avoiding segmentation must be included as a significant factor in the regulatory calculation for realspace sovereigns.

---

[178] At least as reflected in legislation. An interesting aspect of this issue is the apparent tension between the assertive approach taken by the Congress in passing the ACPA, especially the *in rem* provisions, and the more "hands off" version followed by the Clinton Administration. See, e.g., White Paper, Memorandum of Understanding.

[179] A claim, we suggest, that can be seen most expansively in the ACPA's *in rem* proceedings. *See* 15 U.S.C. § 1125(d).

[180] See supra notes __ and accompanying text.

[181] Measurement of the commercial activity online is, of course, subject to various problems. But under any set of criteria, the United States would seem to be doing quite well. The top ten most visited "web properties" worldwide are all US companies. *See* Neilsen/Netratings, *Top 25 Web Properties (October 14, 2001)* at http://pm.netratings.com/nnpm/owa/NRpublicreports.toppropertiesweekly. The top ten electronic commerce sites (measured in terms of customer sales) is entirely American. *See NextCard Ecommerce Index* (September 2001) at http://www.nextcard.com/Indexes/sept_index_movers10_05_01.html. The top ten ecommerce businesses in terms of market share are entirely American. *See Nearly Half of All Americans Buy Online, According to Nielsen//NetRatings and Harris Interactive; $3.5 Billion Spent Online in March, Jumping 36 Percent in Past Year*, BUSINESSWIRE, April 24, 2001.

[182] Note here we are setting aside the many (and significant) non-commercial United States' interests in the growth of the internet, including the spread of democratic values, etc.

We can review this broader approach by explicitly considering the harm due to segmentation. Stylizing the situation as a two-player game, we'll assume two binary choices – either to assert prescriptive jurisdiction and regulate, or defer to entities reflective of other sovereigns' concerns.[183] We consider three outcomes. First, if both states defer, the system will remain unsegmented. If only one state regulates (and no other state retaliates by segmenting), the system will remain unsegmented, with some additional benefit flowing to the regulator. And if both states regulate, the system becomes segmented, resulting in a diminishment of value. Figure 2 sets forth this basic situation, with the payoffs as noted.

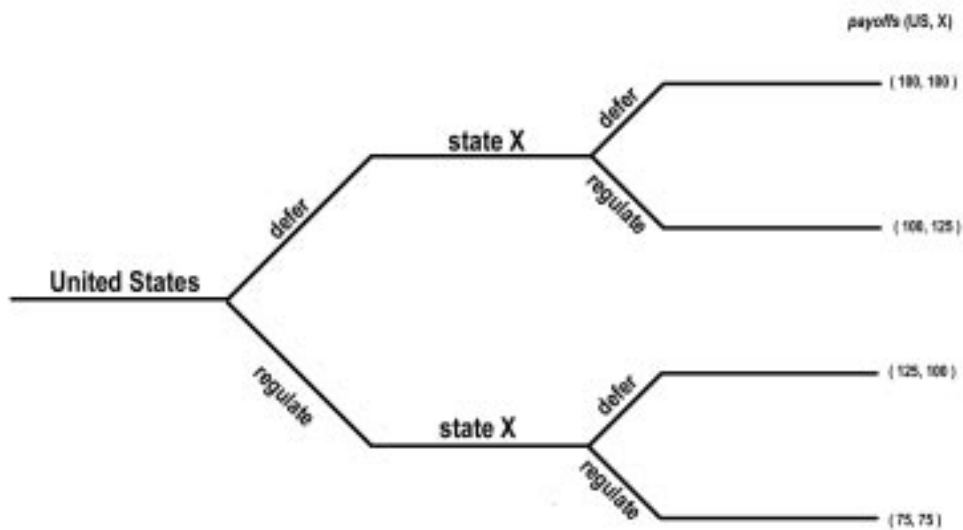

**Figure 2**

Here, the U.S. decision would be to regulate, on the assumption that State X's response would be to defer.[184] Thus, the Figure 1 example might be said to describe the present status quo – where the U.S. has regulated to some degree,[185] and other nations appear to be more deferential.

Yet we suggest that an adjustment in the payoffs could well dramatically alter the nature of this game, as shown in Figure 3.

---

[183] As should be apparent, there are a number of "middle grounds" here that these models fails to adequately capture.

[184] This is a form of a dynamic game with complete information, as state X can easily see the U.S. decision. The sub-game represented by the top branch has expected payoffs of (100, 125), while the bottom branch has payoffs of (125, 100). Accordingly, the U.S. decision will be to regulate. *See* H. Scott Bierman & Luis Fernandez, GAME THEORY WITH ECONOMIC APPLICATIONS 124-135 (1998), Douglas Baird, Robert Gertner & Randal Picker, GAME THEORY AND THE LAW 50-77 (1994), Eric Rasmussen, GAMES & INFORMATION 108 (3rd Ed. 2001).

[185] At least to the extent of the *in rem* provisions of the ACPA.

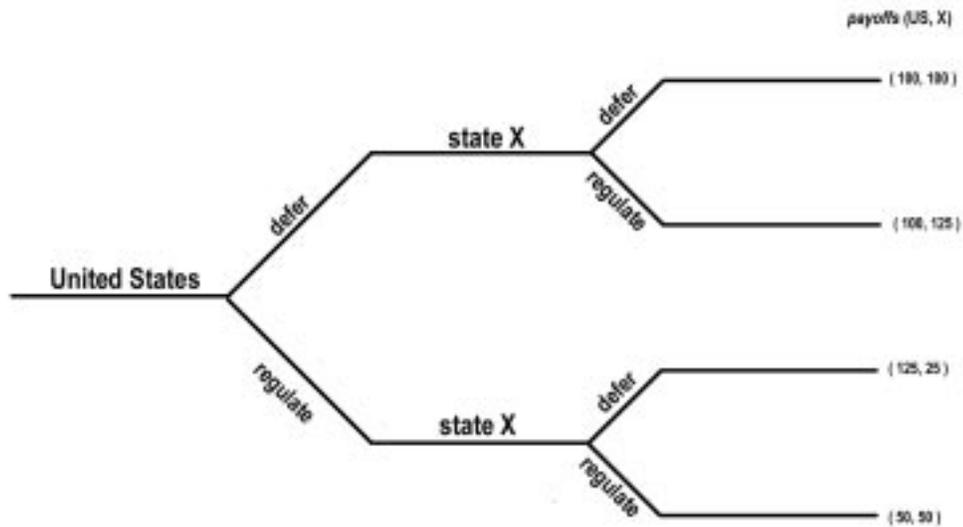

**Figure 3**

Here, we've adjusted the payoff that State X perceives if the United States regulates and it does not. This reduction in payoff could be due to variety of factors – a few of which we've noted above.[186] Under these new conditions, State X will choose to respond to the U.S. regulation with regulation, and the domain name system will become segmented. Accordingly, the best U.S. decision is to defer.

Figure 4 notes one final example, where the U.S. payoff to State X regulation is reduced as well.

---

[186] See supra notes __ and accompanying text.

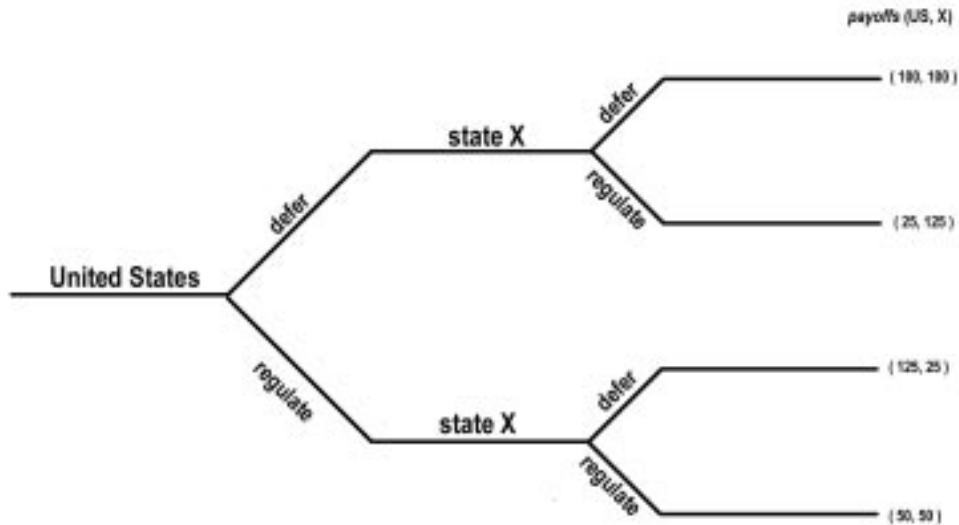

**Figure 4**

Here again, the best U.S. decision is to regulate. That such a decision yields the sub optimal condition of segmentation reveals the nature of this game as a form of the prisoner's dilemma.[187]

(Note of course that a still better view of this problem is as a repeated game – that is, each sovereign has many opportunities to decide whether to regulate or defer. Thus, if the United States considered the assertion of regulation by another sovereign to be more harmful to its interests, it could take the next opportunity to respond by regulating – and thereby further confirm the sub-optimal nature of this situation.)

Towards Coordination

We suggest that the concerns we've outlined above – that the mapping of the logical control over the domain names system onto geographic space lends significant potential for realspace sovereigns to exert regulatory influence on the system, and that the increasing awareness of this power and the availability of alternative root server systems – describe a substantial danger that the present domain name system will become segmented.

And yet, as with most prisoner's dilemmas, coordination provides a solution.[188] In particular, we think that our analysis here establishes the strong interests that realspace sovereigns (and especially the United States) have to coordinate their regulatory behavior – with an eye to avoiding segmentation.

---

[187] The expected payoffs for the top branch are (25, 125), while the expected payoffs for the bottom branch are (50, 50).

[188] See, e.g., Bierman & Fernandez, supra note __ (noting the difference in outcomes between cooperative games and noncooperative games); Baird et al, supra

Such coordination will invariably require greater deference to non-territorial domain name regulatory bodies. The paradigmatic example of this, of course, is ICANN.[189] As many have observed, there are a number of substantial problems with the present form of ICANN;[190] advocating international coordination does not, we think, necessarily require endorsement of the present policies and procedures established by ICANN, or the way in which that organization has been developed.

As many have noted, the challenge of integrating sovereign interests into a coherent international regulatory framework is considerable. We will not revisit that task here. Instead, having noted in particular the perverse incentives created by the current United States regulatory regime, and the strong interests the United States has in avoiding segmentation, we offer a few observations and suggestions for future consideration.

1. We firmly believe that the *in rem* provisions of the ACPA are misguided and should be repealed or substantially revised. As we noted above, they are of only limited value,[191] and appear to serve primarily as a particularly obnoxious example of expansive U.S. claims to regulate domain names.[192]

2. The United States government (and other realspace sovereigns) should take a more active role in supporting the development of international domain name policy coordination. As we've argued throughout this article, the present ICANN approach of avoiding any significant government involvement – and instead attempting to build a strictly nongovernmental regulatory authority – fails to grasp the unavoidable involvement of realspace sovereigns in domain name regulation.[193] Such support will include diplomatic efforts as well as concrete actions that lend additional credibility to these organizations.

3. Finally, Congress should consider revising the ACPA to reflect greater deference to the decisions of international regulatory bodies. For example, Congress might consider implementing a requirement that disputing parties seek resolution from the international domain

---

note __ (observing the importance of "binding agreements" between parties facing a prisoner's dilemma).

[189] The Internet Corporation for Assigned Names and Numbers. See supra notes __ for further description and discussion.

[190] See, e.g. supra notes __ and accompanying text.

[191] See supra notes __ and accompanying text.

[192] See supra notes __ and accompanying text.

[193] See supra notes __ and accompanying text.

name regulatory body prior to filing a Federal lawsuit.[194] Or lawsuits in which a decision has been made by the domain name body might be more limited in their scope, in the nature of an appeal process rather than an initial action.[195]

We are under no illusions that the kind of coordination required to effectively regulate the domain names system will be simple or uncontested. Yet if we are to maintain the value of the domain name system as a solution to the searching and selection problem, realspace sovereigns must recognize the urgent importance of coordination and deference, and tailor their regulatory approach accordingly.

## V

It has become commonplace to describe our world (especially the economic world) as "interdependent." Increasingly, the flow of capital as well as goods and services show little respect for traditional sovereign borders; nations unable or unwilling to respond to economic changes can suffer harm at the hands of the global marketplace.

The advent of the Internet as a powerful commercial (and social) medium is likely to present still greater challenges. For the Internet brings new meaning to "interdependent": in a world where geography is fundamental to our understandings of sovereignty, the contested aspects of online "territoriality" mean that regulation might occur everywhere, or even nowhere. In an era when the "effects" of a commercial dispute in cyberspace might be "felt" both everywhere and nowhere, realspace sovereigns have great power to affect the global progress of the Internet. And there is perhaps no nation with as much at stake in this game as the United States, and no nation with as much power to lead the community of nations in determining the 'net's future.

Yet the evidence of the U.S. approach to date has not been altogether heartening. With respect to the regulation of domain names – perhaps the "canary in the coal mine" of global internet regulation – the U.S. appears, via enactment of the ACPA, and especially its self-styled *in rem* provisions, to have acted in an ill-informed manner that may be contrary to its long-term interests. As we argued above, these statutory provisions suffer

---

[194] The analogy here would be to administrative exhaustion principles.

[195] Note that the ICANN UDRP proceedings explicitly allow parties the right to seek relief from a court prior to actions being taken against a domain name. *See* UDRP, ¶ 4(k). This is in the nature of granting an appellate right; we suggest that Congress may want to formalize such an appellate process as part of an effort to recognize the authority of international domain name regulation. *Cf.* Parisi v. Netlearning, Inc., 139 F. Supp. 2d 745, 752-53 (E.D. Va. 2001) (holding that UDRP decisions do not fall within the limited scope of review for arbitration awards provided by the Federal Arbitration Act).

from the double ignominy of being both of little value[196] and inapt.[197]

We do not, however, believe that all is lost, or that the current U.S. approach to domain name regulation will inexorably lead to the segmentation of the domain name system. To the contrary, the mere presence of the ICANN dispute resolution mechanism as well as the absence (to date) of any serious attempts by other sovereigns[198] to assert jurisdiction over domain names demonstrates, we think, that this problem is not unsolvable at an international scale. But any discussion about the future of international regulation of domain names (and, in a larger sense, the Internet) must proceed with full awareness of the essential role that realspace sovereigns play in both the form and the content of any regulatory approach.

---

[196] Useless in the sense that anytime the section authorizing the *in rem* assertion of jurisdiction (over known persons) will be necessary, it will also violate the US Constitution to accept jurisdiction. See supra notes __ and accompanying text.

[197] Inapt in a strategic sense: because it works an aggressive approach to jurisdiction over domain names that creates significant incentives for other countries to regulate, thus raising the possibility of segmenting the domain name system. See supra notes __ and accompanying text.

[198] Except, perhaps, China. See supra note __ and accompanying text.